\documentclass{aa} 
\usepackage{graphicx}
\usepackage{txfonts}
\usepackage[pdfstartview=FitH,      
            breaklinks=true,        
            bookmarksopen=true,     
            bookmarksnumbered=true, 
            ]{hyperref}
\usepackage{aas_macros}
\usepackage{wasysym} 
\usepackage{subfigure}
\usepackage{placeins}
\usepackage{pdflscape}
\usepackage{xcolor}
\usepackage[version=4]{mhchem}
\newcommand{\comment}[1]{}
\usepackage{ulem}
\usepackage{multirow}

\definecolor{dgreen}{rgb}{0.05,0.5,0.1}
\definecolor{burgundy}{rgb}{0.50,0.00,0.13}

\begin{document}

   \title{The Atmospheres of Rocky Exoplanets}

   \subtitle{I. Outgassing of Common Rock and the Stability of Liquid Water}

   \author{O. Herbort\inst{1,2,3}
          \and
          P. Woitke\inst{1,2}
          \and
          Ch. Helling\inst{1,2,4}
          \and
          A. Zerkle\inst{1,3}
          }

   \institute{
Centre for Exoplanet Science, University of St Andrews, North Haugh, St Andrews, KY169SS, UK\\ \email{oh35@st-andrews.ac.uk}
         \and
SUPA, School of Physics \& Astronomy, University of St Andrews, North Haugh, St Andrews, KY169SS, UK
		\and
School of Earth \& Environmental Studies, University of St Andrews, Irvine Building, St Andrews, KY16 9AL, UK
		\and
SRON Netherlands Institute for Space Research, Sorbonnelaan 2, 3584 CA Utrecht, NL
}
   \date{Received 02 09, 2019; accepted 02 03, 2020}

 
  \abstract
   {Little is known about the interaction between atmospheres and crusts of exoplanets so far, but future space missions and ground-based instruments are expected to detect molecular features in the spectra of hot rocky exoplanets.}
   {We aim to understand the composition of the gas in an exoplanet atmosphere which is in equilibrium with a planetary crust.}
   {The molecular composition of the gas above a surface made of a mixture of solid and liquid materials is determined by assuming phase equilibrium for given pressure, temperature and element abundances. We study total element abundances that represent different parts of the Earth's crust (Continental Crust, Bulk Silicate Earth, Mid Oceanic Ridge Basalt), CI chondrites and abundances measured in polluted white dwarfs.}
   {For temperatures between ${\sim}\,600\,$K and ${\sim}\,3500\,$K, the near-crust atmospheres of all considered total element abundances are mainly composed of \ce{H2O}, \ce{CO2}, \ce{SO2} and in some cases of \ce{O2} and \ce{H2}. 
   For temperatures $\lesssim$\,500\,K, only \ce{N2}-rich or \ce{CH4}-rich atmospheres remain.
   For $\gtrsim$\,3500\,K, the atmospheric gas is mainly composed of atoms (O, Na, Mg, Fe), metal oxides (SiO, NaO, MgO, CaO, AlO, FeO) and some metal hydroxides (KOH, NaOH).
   The inclusion of phyllosilicates as potential condensed species is crucial for lower temperatures, as they can remove water from the gas phase below about 700\,K and inhibit the presence of liquid water.
   } 
   {Measurements of the atmospheric composition could, in principle, characterise the rock composition of exoplanet crusts.
   \ce{H2O}, \ce{O2} and \ce{CH4} are natural products from the outgassing of different kinds of rocks that had time to equilibrate.
   These are discussed as biomarkers, but do emerge naturally as result of the thermodynamic interaction between crust and atmosphere.
   Only the simultaneous detection of all three molecules might be a sufficient biosignature, as it is inconsistent with chemical equilibrium.}

   \maketitle
%
\section{Introduction}\label{sec:Intro}
After the first detections of exoplanets around the pulsar PSR B1257+12 by \citet{1992Natur.355..145W}, and the main sequence star 51Peg by \citet{1995Natur.378..355M}, more than 4000 exoplanet detections have been confirmed to date\footnote{\url{http://exoplanet.eu/catalog/}}. Until recently, the overwhelming  majority of the exoplanets found were gaseous giant planets in close orbits to their host stars.

In the last years, it became possible to detect exoplanets of rocky composition in the habitable zones of their host star, mostly M dwarfs.
Notable detections being the planets around our closest star Proxima Centauri \citep{2016Natur.536..437A}, Barnard's star \citep{2018Natur.563..365R} and the seven planet system around Trappist~1 \citep{2017Natur.542..456G}.
Even though these planets are in close orbits to their host stars, their equilibrium temperatures are lower than 500\,K due to the low effective temperature of their host star.
\citet{2016A&A...596A.112T} showed for Proxima Centauri b that liquid water can be present and might  be detectable in the future. 
On the other hand, rocky exoplanets with short orbital periods around solar-type stars such as 55\,Cnc\,e \citep{2004ApJ...614L..81M} or CoRoT-7b \citep{2009A&A...506..287L} have  temperatures on the day side of $T_{\rm eq}\!\sim\!2400$ K \citep{2011Icar..213....1L, 2016MNRAS.455.2018D}.
This temperature is high enough to melt the surface of the planet on the dayside, while the nightside remains solid \citep[magma pond][]{2016ApJ...828...80K}.
During the formation of terrestrial planets the surface can melt and cause a global magma ocean with a mass dependent depth \citep{2012AREPS..40..113E}.
For short orbital periods and host stars with extensive magnetic fields, this magma ocean can prevail due to inductive heating \citep{2017NatAs...1..878K}.

To date, it is impossible to determine whether the surface material has been processed by plate tectonics or by radiation and stellar winds on an exposed surface only.
For planets with enough heat from e.g. stellar irradiation, inductive heating or tidal forces, an active mantle can prevail that cause active surface processing like volcanism or plate tectonics.
Planetary structure modelling suggests that hot Super-Earths like 55\,Cnc\,e have high atmospheric abundances of refractory elements such as Ca and Al \citep{2019MNRAS.484..712D}.
These atmospheres are shown to allow the formation of mineral clouds \citep{2017MNRAS.472..447M}.

The number of rocky exoplanets whose atmosphere has already been spectroscopically analysed is very small.
However, in the near future the number of exoplanet atmospheres in reach for detailed analysis will increase because of spectrographs with high spectral resolution in the near infrared (e.g. CARMENES \citep{2012SPIE.8446E..0RQ}, CRIRES+ \citep{2014SPIE.9147E..19F}), the 30\,m telescopes and space missions like TESS, JWST and PLATO.
These instruments will allow the analysis of the atmospheric composition of nearest rocky exoplanets with unprecedented precision.

Current studies have revealed a large diversity in the composition of rocky exoplanets, see \url{http://exoplanet.eu} and Fig.~1 in \citet{2017ARA&A..55..433K}. It seems reasonable to expect a comparably large diversity with respect to their atmospheres \citep{2015ExA....40..449L}. The study of exoplanet atmospheres has so far been focused on gas giants (\citealt{2008A&A...492..585D, 2010Natur.465.1049S, 2018Natur.557..526N, 2018ApJ...855L..30A, 2018A&A...620A..97S})  due to observational limitations. \cite{2014Natur.505...69K} observed the super-Earth GJ\,1214\,b and showed that the analysis of the atmosphere's chemical composition is frustrated by clouds.  Atmospheres of exoplanets have been observed to be generally  affected  by cloud formation \citep[for a recent review see][]{2019AREPS..47..583H}.

In what follows, we investigate the atmospheric gas that forms above a surface ('near-crust atmosphere') made of a mixture of solid and liquid materials under certain thermodynamic conditions and for different sets of element abundances. 
We assume thermo-chemical equilibrium for the molecules in the gas phase and phase equilibrium for the condensates in contact with that gas.
Our approach is to study the atmospheric gas in contact with rocky planet's crust and derive the gas composition immediately above the crust. \cite{2012ApJ...755...41S} have presented a similar approach, and \cite{2019MNRAS.482.2893M} used the thermo-chemical equilibrium code TEA (\citealt{2016ApJS..225....4B}) to discuss the atmospheric gas composition of the potential magma-ocean dayside of 55\,Cnc\,e.
All studies assume that the near-crust atmospheric gas had enough time to reach local thermodynamic equilibrium (LTE) with the planetary crust.

We apply the equilibrium code {\sc{GGchem}} \citep{2018A&A...614A...1W} which enables us to calculate the thermo-chemical equilibrium chemistry for gases and condensates in phase equilibrium to temperatures as low as $100\,$K.
{\sc{GGchem}} has been benchmarked against TEA by \cite{2018A&A...614A...1W}.
{\sc{GGchem}} also allows us to investigate  the stability of liquid and solid water at low temperatures. 
\cite{2018A&A...614A...1W} have shown that phyllosilicates become stable below about $500-700$\,K in a solar composition gas in phase equilibrium at $p\!=\!1\,$bar, which then in fact interferes with the stability of liquid water.
Such clay materials can form on the timescales of days on Earth \citep[e.g.][and references therein]{velde2013origin}.
We note that such phyllosilicate materials form already in planet-forming disks \citep{Thi2018} and may contribute substantially to the water delivery to Earth.

We benchmark our code against earlier work by \cite{1990ApJS...72..417S} and compare our results with \cite{2012ApJ...755...41S}.
Different planets in different stellar environments and in different evolutionary states will also differ in their elemental composition which is crucial for the composition of the atmosphere.
We study the effect of different sets of total element abundances. 

We utilize total element abundances informed by geological studies on Earth, Continental Crust (CC), Bulk Silicate Earth (BSE), and Mid Oceanic Ridge Basalt (MORB), and those informed by astronomical studies, solar abundances, CI Chondrites and abundances deduced from  Polluted White Dwarf observations (PWD).
A wide range of pressures (0.001\,bar to 100\,bar) and temperatures (100\,K to 5000\,K) is considered.
Kinetic effects and geological processes such as plate tectonics and volcanism are not included in our model, and we assume that our system had always enough time to reach the thermo-chemical equilibrium after it may have been affected by momentary disequilibrium events like volcanism etc.

In Sect.~\ref{sec:atmosphere_model} we briefly describe the {\sc{GGchem}} code that is used in this work.
A detailed comparison to previously published results is provided in Sect.~\ref{sec:comparison} for element abundances of CC and BSE.
Sect.~\ref{sec:starting_condition} presents additional results for MORB, CI chondrite and PWD abundances.
In Sect.~\ref{sec:abundance_var}, the stability of liquid water is studied by modifying the BSE composition to see by how much the content of gaseous atmospheric water needs to increase for liquid water to be present eventually.
The effect of the atmospheric pressure on the thermo-chemical equilibrium is discussed in Sect.~\ref{sec:pressure}.
Sect.~\ref{sec:timescales} focuses on the estimation of timescales at which the chemical equilibrium is reached.
In Sect.~\ref{sec:discussion} we conclude with a discussion.
In the appendix a glossary of structural formula for some condensed species is provided (Tab. \ref{tab:glossary}).

\section{Method: Phase equilibrium with {\sc{GGchem}}}\label{sec:atmosphere_model}
We use the thermo-chemical equilibrium code {\sc GGchem} \citep{2018A&A...614A...1W}.
Based on the principle of minimisation of the total Gibbs free energy the chemical equilibrium for the molecules in the gas phase and phase equilibrium for the condensates is solved.
A short summary is provided here.
For more details see \citep{2018A&A...614A...1W}.
\begin{figure}
\centering
\includegraphics[width=88mm] {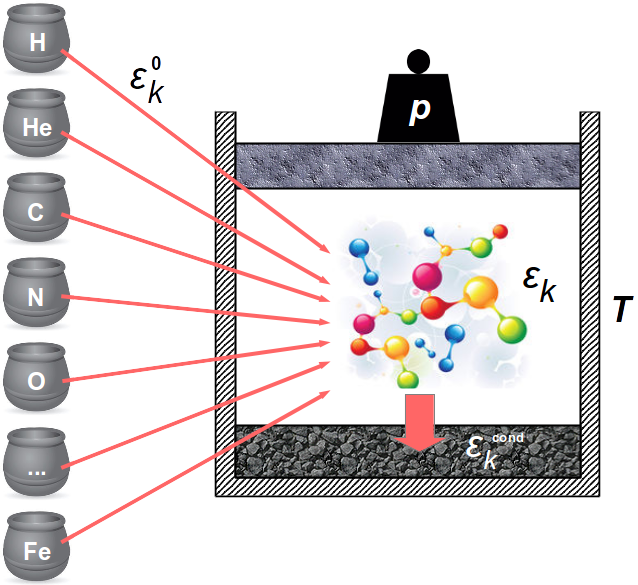}
\caption{The principle modelling procedure of {\sc{GGchem}}: Elements are selected and included with given total abundances $\epsilon^0_k$.
{\sc GGchem} calculates the equilibrium gas phase and condensate abundances at given gas pressure $p$ and temperature $T$, providing the element abundances contained in the condensates ($\epsilon^{\rm cond}_k$) and the remaining gas phase ($\epsilon_k$).}
\label{fig:GGchem}
\vspace*{-2mm}
\end{figure}

Figure~\ref{fig:GGchem} shows the basic procedure: 
A set of elements ${k=1,...,K}$ is selected, their abundances with respect to hydrogen, $\epsilon_k^0$, are henceforth called the {\it total element abundances}. 
Based on gas pressure $p$, gas temperature $T$ and $\epsilon_k^0$, {\sc GGchem} determines the stable condensates, calculates their abundances (which results in the abundance of condensed elements $\epsilon_k^\mathrm{cond}$) and calculates the ion, atom and molecular concentrations in the gas (which results in the element abundances left in the gas phase $\epsilon_k$).
Condensates ${j\!=\!1,...,J}$ are considered stable and present if their supersaturation ratios are unity $S_{\!j}=1$, whereas all other condensates are unstable and not present, i.e. $S_{\!j}<1$.
The conservation of elements $\epsilon_k^0 = \epsilon_k + \epsilon_k^\mathrm{cond}$ is obeyed for all elements $k$.
The resulting abundances of all gas and condensed phase species obey the conditions of chemical and phase equilibrium.

In many astrophysical objects, such as AGB star winds, brown dwarf atmospheres or protoplanetary discs, the majority of elements remains in the gas phase, i.e.\ ${\sum\epsilon_k^0 \approx \sum\epsilon_k \gg \sum\epsilon_k}$.
However, for most applications considered in this paper, such as a gas in contact with a hot planetary crust, only a small fraction of the elements ($10^{-1}...10^{-5}$) will actually remain in the gas phase, depending on the temperature, i.e. ${\sum\epsilon_k^0 \approx \sum\epsilon_k^{\rm cond} \gg \sum\epsilon_k^\mathrm{cond}}$.
Therefore, the vapour pressure of stable condensates is a major component for the composition of the gas.

{\sc GGchem} combines different thermo-chemical data sources. For the molecular equilibrium constants, $k_p(T)$, we use \cite{Stock} and \cite{2016A&A...588A..96B}, complemented by some new fits to the NIST/JANAF database \citep{1982JPhCS..11..695C, 1986jtt..book.....C, chase_mono9}.
The condensed phase data $\Delta G_\mathrm{f}^0(T)$ is taken from {\sc SUPCRTBL} \citep{2016CG.....90...97Z} and {\sc NIST/JANAF}.
These datasets allow us to calculate the concentrations of all atoms, ions, molecules and condensed phases in chemical equilibrium for a mixture of up to 41 elements (hydrogen to zirconium, and tungsten).
The actual number of gas phase species and condensates depends on the selection of elements and will be stated in the corresponding sections for the different calculations.

Phase equilibrium models can generally provide only a very simplified and limited approach to describe the occurrence of condensates in gases, and the nature of gases above solid surfaces, especially when low temperatures are considered, where both the outgassing and the deposition rates are small.
No kinetic rates are considered, and the relaxation timescale toward chemical and phase equilibrium cannot easily be discussed.
For example, the atmosphere of a rocky planet is in contact with the crust partly on the hot dayside and partly on the cold nightside.
Other processes are not considered, such as cloud formation, photo-dissociation and cosmic ray induced processes, volcanism, etc.
Still, equilibrium models can provide a first understanding and can be used to inform more ambitious kinetic condensation models, for example, about the choice of condensates.

\section{Comparison to previous phase equilibrium models} \label{sec:comparison}
We compare the phase-equilibrium results obtained with {\sc GGchem} to previous equilibrium condensation models published by \cite{1990ApJS...72..417S} and \cite{2012ApJ...755...41S} in the following Sects.~\ref{ssec:sharphuebner} and \ref{ssec:schaefer}, respectively.
The paper by \cite{1990ApJS...72..417S} includes sufficiently detailed information about the selection of molecules and condensates, and their thermo-chemical data, that allow us to benchmark our results.
For \cite{2012ApJ...755...41S}, we do not exactly know their choice of molecules and condensates, nor their thermo-chemical data applied, so we broadly compare our results,
and identify the differences between the two models.

\subsection{Sharp \& Huebner}\label{ssec:sharphuebner}
\begin{figure}
\vspace*{-2mm}
\hspace*{-3mm}
\includegraphics[width = 93mm] {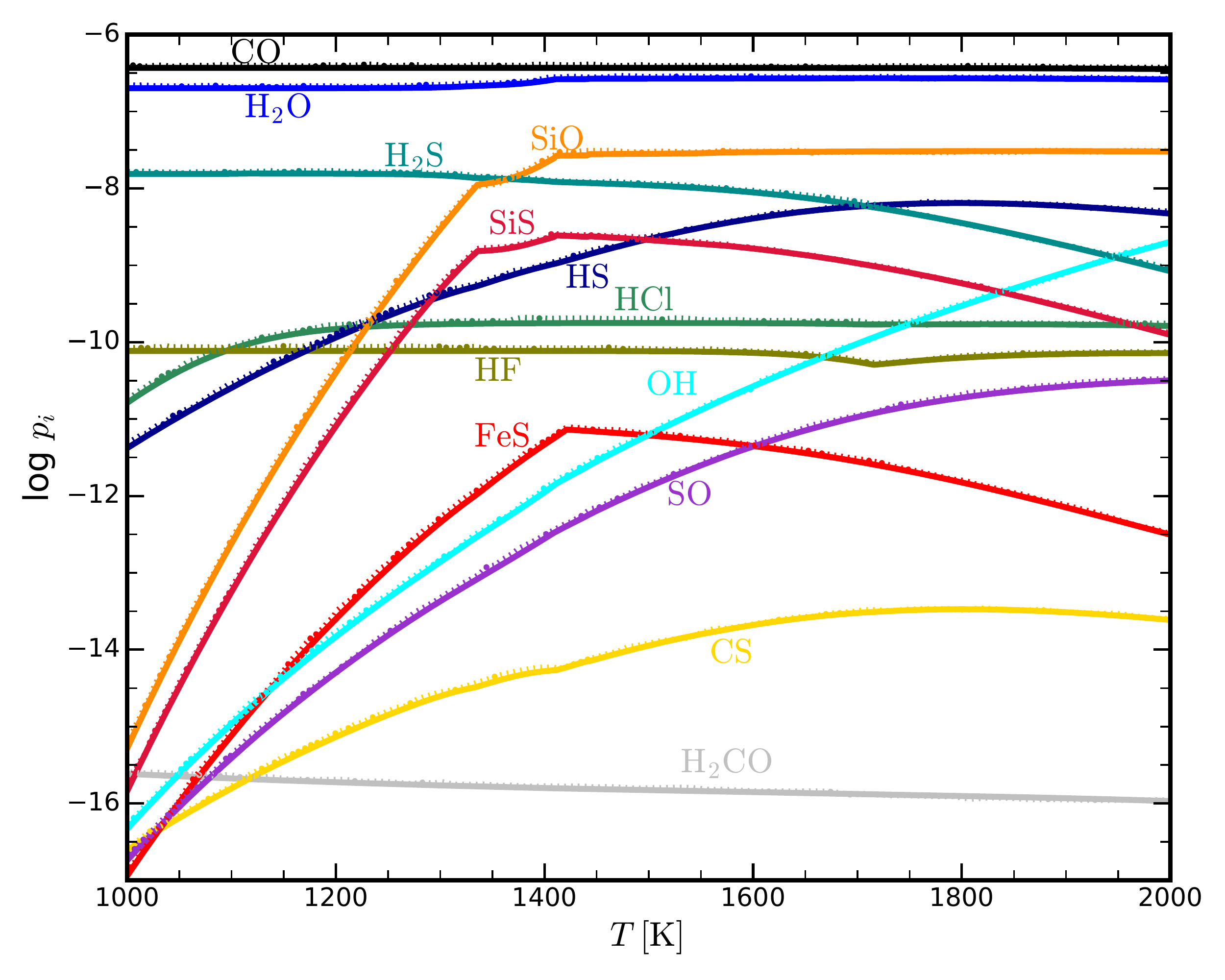}\\[-2ex]
\hspace*{-3mm}
\includegraphics[width = 93mm] {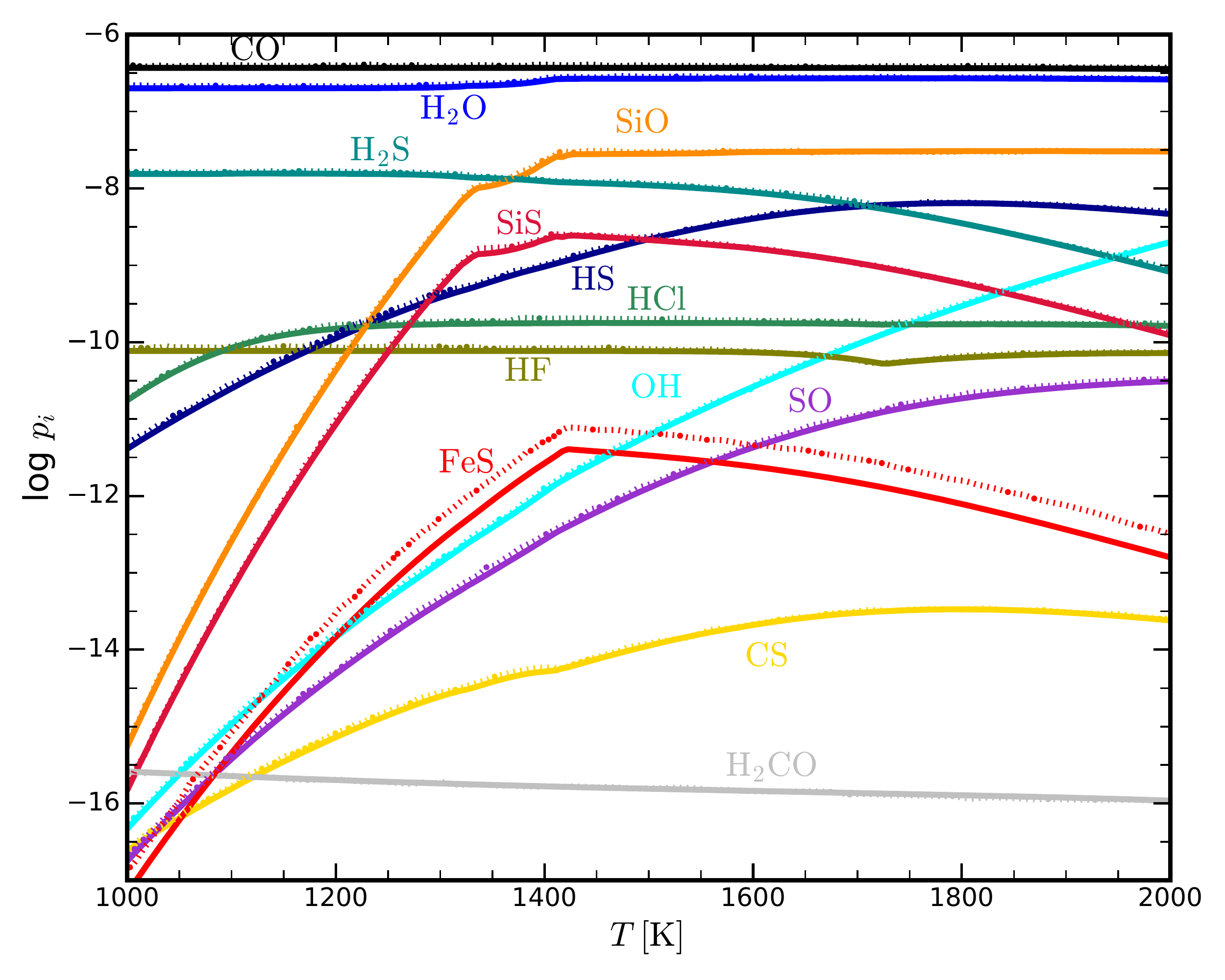}\\[-4ex]
\caption{Comparison of molecular partial pressures computed by {\sc GGchem} (solid lines) to the results by \citet[][dotted lines]{1990ApJS...72..417S}. Calculations are based on the solar element abundances listed in \cite{1990ApJS...72..417S} for a constant pressure of $p\!=\!0.5\,$mbar.
The shown elements are chosen to match Fig.~1 in \citet{1990ApJS...72..417S}.
{\bf Upper panel:} {\sc GGchem} only uses the molecules and condensates selected by \cite{1990ApJS...72..417S} and applies their thermo-chemical data, {\bf Lower panel:} {\sc GGchem} uses all molecules and condensates found in its own database and uses its own thermo-chemical data.
The agreement is very good. The only visible difference is in \ce{FeS}, which is slightly depleted in the lower panel.}
\label{fig:comp_sharp}
\vspace*{-3mm}
\end{figure}

\citet{1990ApJS...72..417S} used a numerical method that directly minimises the total Gibbs free energy of gas phase and condensed phase species. {\sc{GGchem}} uses a different numerical method, for details see \citet{2018A&A...614A...1W}, so the aim of this section is mainly to verify our numerical approach.
Therefore the chosen elements, element abundances (Tab.~\ref{tab:abundances}) and the thermo-chemical data for the gas and condensate species are exactly as described in \cite{1990ApJS...72..417S}.
Thus both codes use the same 18 elements (H, C, N, O, F, Na, Mg, Al, Si, P, S, Cl, K, Ca, Sc, Ti, V, Cr, Mn, Fe, Ni, Cu and Zr) with 165 gas species and 62 solid condensed species.
As in \cite{1990ApJS...72..417S}, the pressure is set to $0.5\,$mbar and the temperature ranges from $1000\,$K to $2000\,$K.

In Fig.~\ref{fig:comp_sharp}, the resulting {\sc GGchem} molecular partial pressures are shown, with the Sharp \& Huebner results over plotted as dashed lines.
The graphs from Sharp \& Huebner's figure~1 have been digitised from an electronic version of their paper. Within the precision of that digitising process, our results are identical.
In addition, the "appearance and disappearance temperatures" of the condensates listed in Table~3 of \citet{1990ApJS...72..417S} agree within $\pm 1\,$K with our results. This demonstrates that our numerical method produces equivalent results for both, condensates and molecules.

In the lower panel of Fig.~\ref{fig:comp_sharp}, we have applied the full {\sc{GGchem}} dataset for molecules and condensed species to the same selection of elements and their abundances, resulting in 388 gas-phase and 204 condensed species, 35 of them being liquids.
The results are still very close, with \ce{FeS} being the only molecule with relevant concentrations that shows a lower concentration compared to the results by \cite{1990ApJS...72..417S}.
This is caused by the additional occurrence of \ce{FeH} as a gas phase species, which was not included in \cite{1990ApJS...72..417S}.
Additional differences in the selection of molecules and condensates are as follows: 
All gas species and condensates in \cite{1990ApJS...72..417S} are also considered in the full {\sc GGchem} model.
We find that \ce{SiO}[s], \ce{ZrSiO4}[s], \ce{CaTiSiO5}[s] and \ce{MgCr2O4}[s] become stable condensates, while \ce{MgTi2O5}[s] and \ce{Cr2O3}[s] are not becoming stable.
In the gas phase of these two models, the major difference is that our \ce{CrH} and \ce{CuH} concentrations differ by more than one order of magnitude.
This is caused by the stability of the additional gas phase species \ce{NiCl}, \ce{TiF3}, \ce{TiOF2}, \ce{TiOCl2}, \ce{TiOCl}, \ce{AlF2O}, \ce{MnF}, \ce{MnCl}, \ce{ZrF3}, \ce{ZrF4}, \ce{CaCl}, \ce{P4O6}, \ce{PO2}, \ce{PCH}, \ce{FeH}, \ce{CrS} and \ce{TiH}, which are all not included in \citet{1990ApJS...72..417S}.

\subsection{Schaefer et al.}\label{ssec:schaefer}
Two sets of total element abundances were considered by \cite{2012ApJ...755...41S} for the calculation of the atmospheric composition above a rocky planet: Continental Crust (CC) and Bulk Silicate Earth (BSE), based on \citet{HANSWEDEPOHL19951217} and \citet{1993Icar..105....1K}, respectively.
The corresponding element abundances are listed in Table~\ref{tab:abundances}. The following 18 elements are selected: H, C, N, O, F, Na, Mg, Al, Si, P, S, Cl, K, Ca, Ti, Cr, Mn and Fe, together with their respective single ions. The pressure is kept constant at 100\,bar.

The model used in \cite{2012ApJ...755...41S} is based on \mbox{IVTANTHERMO} \citep{BELOV1999173} with a database including data from \citet{1995BUSGS2131..461R, HollandPowell2011}.
The phyllosilicates present in their dataset were excluded as their focus was on higher temperatures.
In order to better compare our model to \cite{2012ApJ...755...41S}, we
computed two models with {\sc{GGchem}}, one with and one without phyllosilicates.
For the selected elements, {\sc GGchem} finds 471 gas species in its database and 188 condensates, where 30 of those are liquids and 39 are phyllosilicates.

Phyllosilicates should be an integral part of equilibrium condensation models as they are known in geology to form effectively in a wide temperature range on relatively short timescales as condensates directly from the gas phase, but as alterations in silicate rocks exposed to water vapour \citep[e.g.][and references therein]{velde2013origin}.
The effective formation of phyllosilicates is underlined by spectrometric evidence of phyllosilicates on Mars' surface \citep{poulet2005}.
Based on laboratory analysis of different chondrite materials, using backscattered electron micrography, \citet{bischoff1998} concluded for many carbonaceous chondrites that the aqueous alteration of parent bodies is a fundamental process in their evolution and even argued for pre-accretionary aqueous alteration of distinct components in carbonaceous chondrites.
In the cold midplanes of protoplanetary discs, \cite{DAngelo2019} showed that the hydration of forsterite surfaces by water vapour adsorption should occur within the lifetime of the solar nebula at densities $\sim\!10^8\rm\,cm^{-3}$ and temperatures lower than 500\,K, providing between 0.5 and 10 times the amount of water in Earth's oceans to the bulk composition of Earth, depending on grain size. 
\citet{Thi2018} argued that the chemisorption sites for OH and $\rm H_2O$ molecules in the silicate cores become occupied at temperatures between 250\,K and 700\,K on timescales shorter than $10^5\rm\,yrs$ for 1\,mm grains at gas densities of $10^8\rm\,cm^{-3}$.

\begin{figure}
\hspace*{-4mm}
\includegraphics[width = 94mm,page=1] {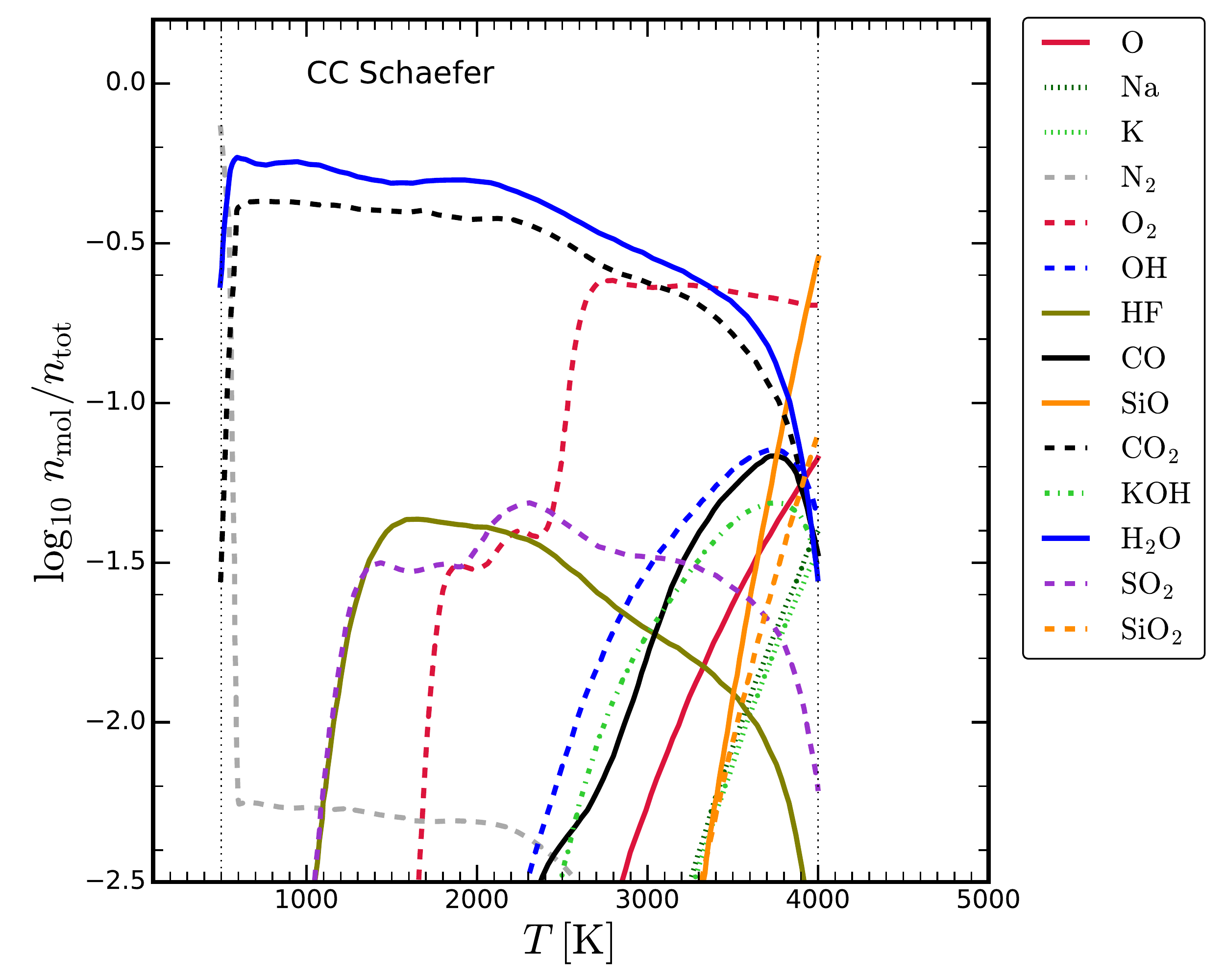}\\[-1ex]
\hspace*{-4mm}
\includegraphics[width = 94mm,page=3] {./graphs/comparison.pdf}\\[-1ex]
\hspace*{-4mm}
\includegraphics[width = 94mm,page=2] {./graphs/comparison.pdf}\\[-3.7ex]
\caption{Comparing the molecular concentrations ($n_\mathrm{mol}/n_\mathrm{tot}$) over Continental Crust (CC) at $P\!=\!100\,$bar predicted by {\sc GGchem} with the results obtained by \cite{2012ApJ...755...41S}.  {\bf Top panel:} results by \cite{2012ApJ...755...41S} scanned from their Fig.~1.  {\bf Middle panel:} results from {\sc GGchem} disregarding phyllosilicates. {\bf Bottom panel:} results from {\sc GGchem} for the full dataset. All species with $\log\!n_\mathrm{mol}/n_\mathrm{tot}\!>\! 1.5$ are shown.
The thin, vertical dotted lines indicate the upper and lower limits of the models.
For the {\sc GGchem} models this means that all elements are condensed below the corresponding temperature.}
\label{fig:comparison_CC}
\vspace*{-5mm}
\end{figure}

\paragraph{a) Continental Crust}
is a proxy of an \ce{SiO2}-rich (felsic) elemental composition which on Earth results from plate tectonics, see e.g.\ \citet{1995RvGeo..33..241T} for a review.
The crust of Earth differentiated into two different parts, continental and oceanic crust. The oceanic crust is geologically younger and consists mainly of basalt (\ce{SiO2}-poor igneous rock) whereas the continental crust is much older, less dense, and consists mainly of {\sl granite}, which is a composite material mostly made of \ce{SiO2} ({\sl quartz}) and {\sl feldspar}.
Plate tectonics on Earth strongly effected the differentiation into the different crust types.
Whether other planets have a similar bulk crust composition remains unclear, but planets with Earth-like plate tectonics can be expected to also have a felsic composition, at least in parts of their crusts.
Furthermore bulk compositions with high \ce{SiO2} abundances can result in overall felsic compositions.
On Venus, sprectrometer measurements from the Galileo mission suggest that the highlands are felsic in composition \citep[e.g.][]{2008JGRE..113.0B24H}, while for Mars, the buried rock underneath the dominant basaltic surface might be of felsic composition \citep[e.g.][]{2017Icar..288..265C}.
The overall composition of the CC is controversial and leads to different proposed compositions.
For further reading, see e.g.\ \citet{1995RvGeo..33..241T}, \citet{2003TrGeo...3....1R} and \citet{Greber1271}.
In order to allow a comparison to \citet{2012ApJ...755...41S} we use their composition taken from \citet{HANSWEDEPOHL19951217}.

Overall, the results of the model by \citet{2012ApJ...755...41S} and our {\sc GGchem} model without phyllosilicates are similar for CC abundances, both showing \ce{H2O} and \ce{CO2} as the major gas species between about 570\,K and 3500\,K, with constant particle ratio \ce{H2O}:\ce{CO2}$=$1.35:1, which is a direct consequence of the assumed total element abundances, since neither hydrogen nor carbon can condense at these temperatures.

At higher temperatures $\ga\!3500\,$K, \ce{O2} and \ce{SiO} become the most abundant species, followed by \ce{CO}, \ce{OH}, metal oxides such as \ce{SiO}, \ce{SiO2}, \ce{NaO} and \ce{AlO}, and hydroxides such as \ce{NaOH} and \ce{KOH}, and by the atoms \ce{K} and \ce{Na}. 
Liquid \ce{Al2O3}[l] is found to be stable for temperatures up to 5150\,K at 100\,bar in our model, where it is the first stable condensate. Other liquid metal oxides are stable at high temperatures as well, in particular \ce{MgAl2O4}[l], ce{CaO}[l], \ce{FeO}[l], \ce{SiO2}[l], \ce{K2SiO3}[l], \ce{Na2SiO3}[l], with the last liquid solidifying at 1696\,K, the melting point of \ce{SiO2}.

At lower temperatures, {\sc GGchem} identifies the following solid compounds as abundant stable condensates (with 
$\max\,\{\,\log_{10}(n_{\rm solid}/n_{\langle\rm{Si}\rangle})\,\}\!\!>\!\!-2$): 
\ce{SiO2}[s] ({\sl quartz}, by far the most abundant), 
the three major components of {\sl feldspar}: \ce{CaAl2Si2O8}[s] ({\sl anorthite}), \ce{NaAlSi3O8}[s] ({\sl albite}) and \ce{KAlSi3O8}[s] ({\sl microcline}),
\ce{MgSiO3}[s] ({\sl enstatite}), 
\ce{Mg2SiO4}[s] ({\sl fosterite}),
\ce{CaMgSi2O6}[s] ({\sl diopside}), 
\ce{KAlSi2O6}[s] ({\sl leucite}), 
\ce{Fe2O3}[s] ({\sl hematite}), 
\ce{Fe3O4}[s] ({\sl magnetite}), 
\ce{CaSiO3}[s] ({\sl wollastonite}) and
\ce{Ca2SiO4}[s] ({\sl larnite}).

At temperatures $\la\!570\,$K, \ce{H2O} gas becomes liquid, and carbon becomes thermally stable as \ce{CaMgC2O6}[s] ({\sl dolomite}), and so \ce{N2} remains the only abundant gas species.
Eventually, at 360\,K, nitrogen is found to be stable in form of \ce{NH4Cl} ({\sl ammonium chloride}) and there is no physical solution anymore to produce a $p\!=\!100\,$bar gas as all selected elements condense.
Such atmospheres would mainly consist of noble gases, which are not included here.

Some deviations between the two models are found for \ce{O2}, \ce{SO2} and \ce{HF}. The results by \citet{2012ApJ...755...41S} show two major steps in the \ce{O2} concentration at $\sim2600\,$K and at $\sim1800\,$K.
These changes in \ce{O2} abundances in the gas phase are caused by the chemical stability of different Fe species in the corresponding temperature regimes:
\ce{Fe2SiO4}[l] ($T>2500\,$K), \ce{Fe3O4}[l] ($1900\,\mathrm{K}<T<2500\,$K) and \ce{Fe2O3}[l]($T<1900\,$K).
In contrast, in the {\sc GGchem}-model, there is only one transition, at about $2150\,$K, where one of the major liquids FeO[l] solidifies to form \ce{Fe3O4}[s] ({\sl magnetite}), which consumes large amounts of molecular oxygen as
\begin{equation}
  \hfill
  \rm 6\;FeO[l] ~+~ O_2\  
  ~\longleftrightarrow~
  \rm 2\;Fe_3O_4[s] \ .
  \nonumber
  \hfill
\end{equation}
The liquid phases \ce{Fe2SiO4}[l], \ce{Fe3O4}[l] and \ce{Fe2O3}[l] are currently not included in our model.
The thermodynamic data for liquids originates mostly from NIST/JANAF, which does not include liquid phases of \ce{Fe2SiO4}[l], \ce{Fe3O4}[l] and \ce{Fe2O3}[l].
Towards higher temperatures, the \ce{O2} abundance increases further according to its increasing vapour pressure over the mostly liquid condensates, such as \ce{SiO2}[l], \ce{MgSiO3}[l] and \ce{FeO}[l].

Towards lower temperatures, fluorine becomes stable in form of \ce{CaF2}[s] ({\sl fluorite}) around 1500\,K,  reducing \ce{HF} in the gas phase.
At about 1150\,K, sulphur becomes thermally stable in form of \ce{CaSO4}[s] ({\sl anhydrite}) on the expense of \ce{CaMgSi2O6}[s] ({\sl diopside}).
This transition removes sulphur from the gas phase, leading to a fast decrease of the \ce{SO2} concentration below 1150\,K.
Eventually, in our model without phyllosilicates, \ce{H2O}[l] ({\sl liquid water}) is thermally stable for $T\!\la\!570\,$K and leaves behind an \ce{N2}-rich atmosphere.

\begin{figure}[!htbp]
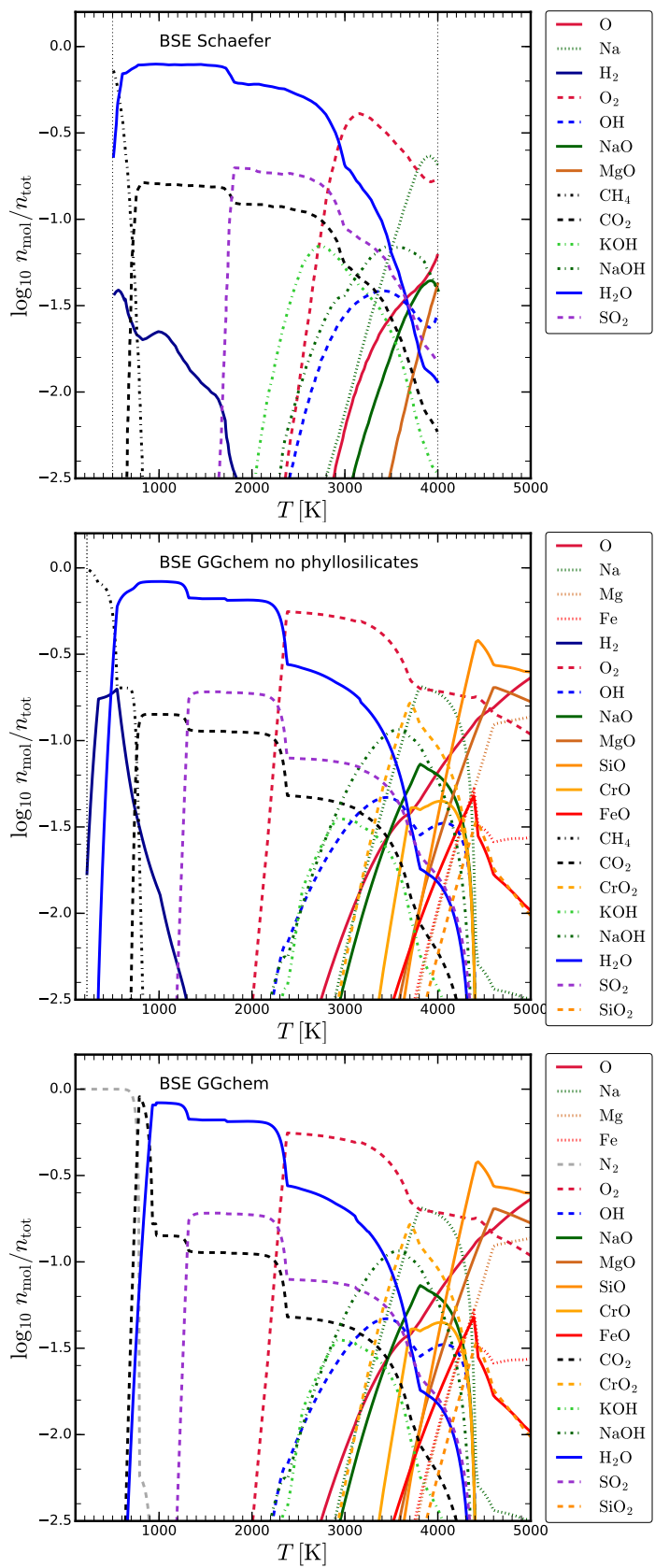

\hspace*{-4mm}
\includegraphics[width = 94mm,page=4] {./graphs/comparison.pdf}\\[-1ex]
\hspace*{-4mm}
\includegraphics[width = 94mm,page=6] {./graphs/comparison.pdf}\\[-1ex]
\hspace*{-4mm}
\includegraphics[width = 94mm,page=5] {./graphs/comparison.pdf}\\[-3.7ex]
\caption{Similar to Fig.\,\ref{fig:comparison_CC} but for Bulk Silicate Earth (BSE) total element abundances.}
\label{fig:comparison_BSE}
\vspace*{-2mm}
\end{figure}

When we do include phyllosilicates in our model (lower panel of Fig.\,\ref{fig:comparison_CC}), the atmospheric composition does not change at higher temperatures, however, at $T$\,$<$\,$720\,$K, phyllosilicates become stable and have a profound influence on both the molecular and solid composition.
We find two phyllosilicates in our CC model, \ce{Mg3Si4O12H2}[s] ({\sl talc}) at $T\!<\!720$\,K, and \ce{Ca2FeAl2Si3O13H}[s] ({\sl epidote}) at $T\!<\!410$\,K. These are very effective in removing water from the gas phase, for example
\begin{equation}
  \hfill
  \rm SiO_2[s] ~+~ 3\;MgSiO_3[s] ~+~ H_2O
  ~\longleftrightarrow~
  \rm Mg_3Si_4O_{12}H_2[s] \ ,
  \nonumber
  \hfill
\end{equation}
thereby inhibiting the stability of liquid water on the surface.
Since hydrogen disappears sooner than carbon from the gas phase in the model including phyllosilicates, there is in fact a narrow temperature interval $\rm 600\,K-720\,K$ where \ce{CO2} is the most abundant molecule.
\paragraph{b) Bulk Silicate Earth}
is an approximation for the composition of the Earth excluding its core. This leads to a composition that is rich in \ce{MgO} and \ce{FeO}-bearing silicates (mafic), but relatively poor in C, N, F, P, S, Cl and K.
For the model comparison we use the same total element abundance as \citet{2012ApJ...755...41S} taken from \citet{1993Icar..105....1K}.

Relevant liquids at high temperatures in our {\sc GGchem}-model are \ce{SiO2}[l], \ce{FeO}[l], \ce{MgSiO3}[l], \ce{Mg2SiO4}[l], \ce{MgAl2O4}[l], \ce{MgO}[l] and \ce{CaO}[l], with the first stable condensate being \ce{MgAl2O4}[l] at 4900\,K. The relevant solid composition at lower temperatures includes 
\ce{Mg2SiO4}[s] ({\sl fosterite}),
\ce{MgSiO3}[s] ({\sl enstatite}),
\ce{FeO}[s] ({\sl ferropericlase}),
\ce{CaMgSi2O6}[s] ({\sl diopside}),
\ce{MgAl2O4}[s] ({\sl spinel}), 
\ce{NaAlSiO4}[s] ({\sl nepheline}),
\ce{Fe2SiO4}[s] ({\sl fayalite}), 
\ce{Fe3O4}[s] ({\sl magnetite}),
\ce{FeAl2O4}[s] ({\sl hercynite}),
\ce{CaAl2Si2O8}[s] ({\sl anorthite}),
\ce{NaAlSi3O8}[s] ({\sl albite}),
\ce{NaAlSi2O6}[s] ({\sl jadeite}),
\ce{MgFe2O4}[s] ({\sl magnetoferrite}),
\ce{Ca3Al2Si3O12}[s] ({\sl grossular}),
\ce{Ca2MgSi2O7}[s] ({\sl \aa kermanite}) and
\ce{Ca2SiO4}[s] ({\sl larnite}).

Concerning the gas phase abundances, both the  \citet{2012ApJ...755...41S} model and our {\sc GGchem}-model show that \ce{H2O} and \ce{O2} are the most abundant gas species in a wide temperature range, but the deviations between the models are more pronounced for BSE abundances.

For high temperatures $\ga\!4200\,$K, both models find the gas phase to be mostly composed of O, Na, Mg, Fe, SiO and MgO, in addition to \ce{O2}. The most obvious deviation is the concentration of \ce{O2}, which dominates the gas phase in our {\sc GGchem}-model between about $2300\,$K and $4200\,$K, whereas it drops already at $3100\,$K in the model by \citet{2012ApJ...755...41S}.
Similar to the model for CC abundances, the disappearance of \ce{O2} in our model for BSE abundances is caused by the solidification of \ce{FeO}[l] around 2300\,K, which consumes oxygen as
\begin{equation}
  \hfill
  \rm Mg_2SiO_4[l] ~+~ 2\;FeO[l] ~+~ \textstyle{\frac{1}{2}}\,O_2
  ~\longleftrightarrow~
  \rm MgSiO_3[l] ~+~ MgFe_2O_4[s] \ .
  \nonumber
  \hfill
\end{equation}
Once \ce{O2} has disappeared, \ce{H2O} becomes the most
abundant and \ce{SO2} the second most abundant species. At $T\!\approx\!1300\,$K, sulphur condenses in form of \ce{FeS}[s] ({\sl troilite}), for example as
\begin{equation}
  \hfill
  \rm 2\;FeO[s] ~+~ 2\;SO_2
  ~\longleftrightarrow~
  \rm 2\;FeS[s] ~+~ 3\;O_2 \ .
  \nonumber
  \hfill
\end{equation}
This removes \ce{SO2} and hence sulphur from the gas phase, making \ce{CO2} the second most abundant gas species after \ce{H2O}.

For $T\!\la\!550\,$K, \ce{H2O} is thermally stable as liquid water \ce{H2O}[l].
Both our {\sc GGchem} model and the \citet{2012ApJ...755...41S} model suggest that, after water is stable as a liquid, the atmosphere becomes rich in \ce{CH4}, with some traces of \ce{H2}. 
At even lower temperatures, the {\sc GGchem} model predicts that \ce{NH4Cl} ({\sl ammonium chloride}) becomes stable at 330\,K and \ce{CH4}[s] ({\sl methane ice}) at 220\,K, below which there is no solution anymore for a 100\,bar atmosphere.

At temperatures below about 1000\,K, the inclusion of phyllosilicates (lower panel in Fig.~\ref{fig:comparison_BSE}) again leads to substantial differences between our model and \citet{2012ApJ...755...41S}.
As the phyllosilicates \ce{KMg3AlSi3O12H2}[s]  ({\sl phlogopite}) and \ce{NaMg3AlSi3O12H2}[s] ({\sl sodaphlogopite}) become stable at temperatures of 970\,K and 930\,K, respectively, the water in the atmosphere is removed.
This removal leads to an earlier domination of \ce{CO2} and subsequently to the formation of solid \ce{C}[s] ({\sl graphite}) at about 800\,K, and the development of an \ce{N2}-rich atmosphere, in contrast to the model without phyllosilicates.
In fact, nitrogen stays in the atmosphere down to 100\,K and does not condense in form of \ce{NH4Cl} as in our CC model.

Our models without phyllosilicates match those of \citet{2012ApJ...755...41S}, which do not include phyllosilicates.
The inclusion of phyllosilicates as condensed species causes strong deviations in the gas phases for temperatures below $\sim1000\,$K.
This underlines the importance of phyllosilicates for the investigation of atmospheres of rocky exoplanets.

\section{Other total element abundances}
\label{sec:starting_condition}

\begin{figure}
  \hspace*{-5mm}
  \includegraphics[width=97mm]{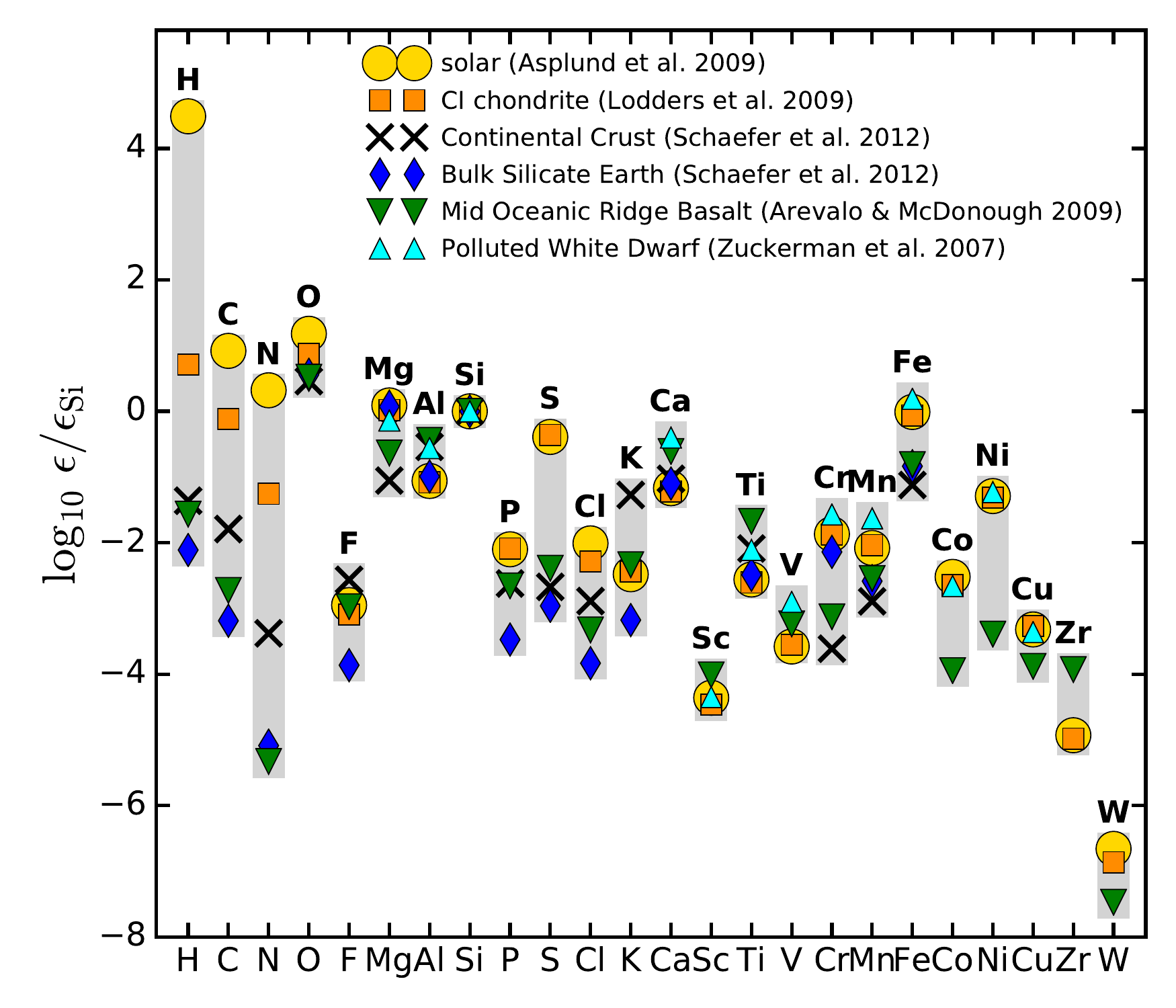}\\*[-4ex]
  \caption{Element abundances relative to silicon (nuclei particle ratios) for different materials and astronomical sources.}
  \label{fig:Abundances}
\end{figure}

One of the major aims and challenges of exoplanet research is to determine the element composition of surfaces and atmospheres, for the purpose of characterisation.
After having studied the theoretical predictions for Continental Crust (CC) and Bulk Silicate Earth (BSE) element abundances in the previous section, where we also have checked our results against previously published results, we will now explore three additional sets of total element abundances for our phase equilibrium models:
(1) Mid Oceanic Ridge Basalt (MORB) elemental abundances,
(2) measured elemental abundances of carboneceous chondrites (CI), and
(3) exoplanet elemental abundances deduced from spectral analyses of polluted white dwarfs.
In Fig.~\ref{fig:Abundances}, we plot these element abundances in units of the silicon abundance, the dominant rock-forming element, and compare them to the element abundances discussed earlier in this paper.

\begin{figure}[!htbp]
\hspace*{-4mm}
\includegraphics[width = 94mm, page=8] {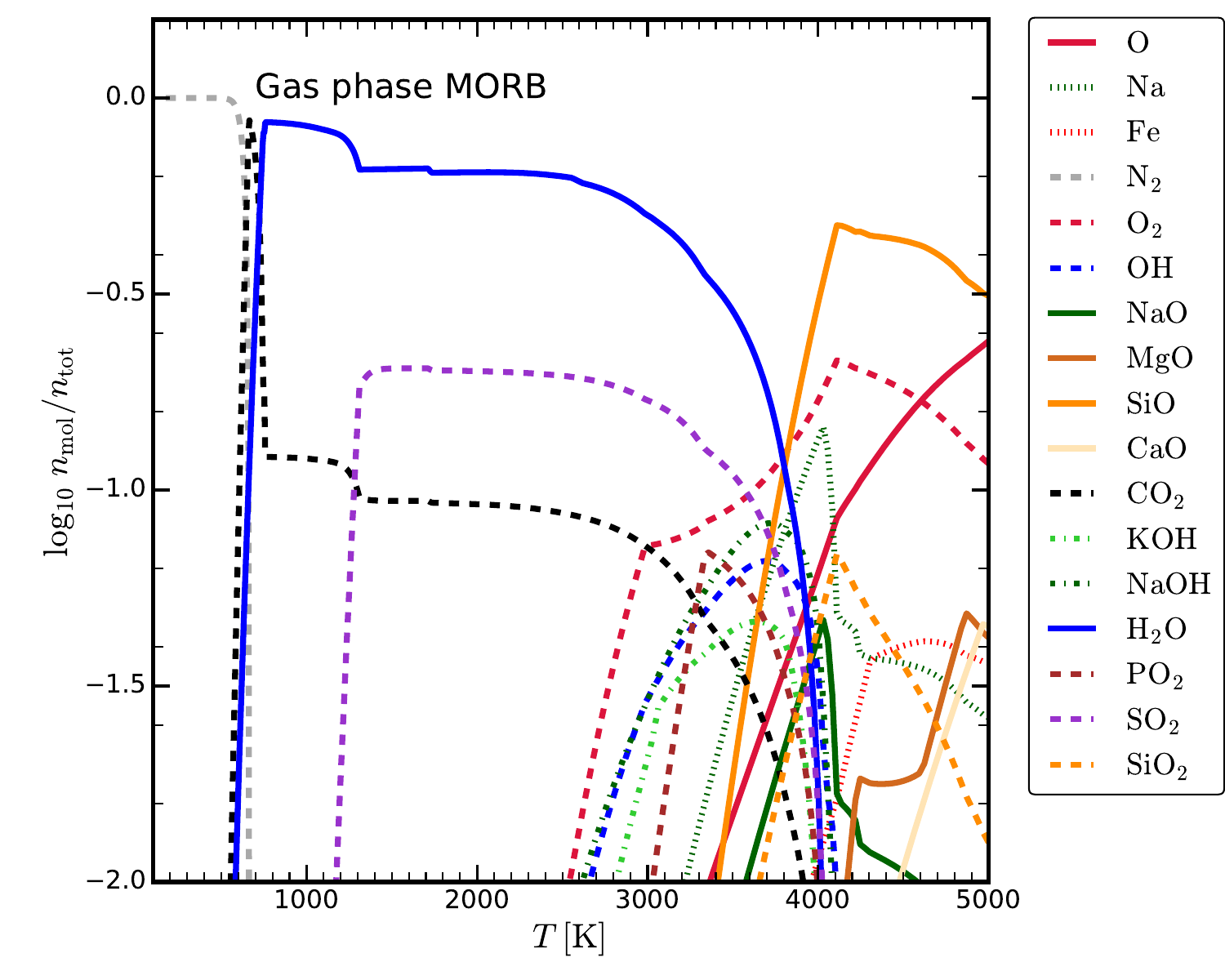}\\[-4ex]
\caption{Molecular concentrations $n_\mathrm{mol}/n_\mathrm{tot}$ for Mid Oceanic Ridge Basalt (MORB) element abundances as function of temperature for a constant pressure of $p\!=\!100\,$bar.
Phyllosilicates are included in this model. All species with maximum log concentration $>\!-1.4$ are shown.
}
\label{fig:GGchem-MORB}
\end{figure}

\subsection{Mid Oceanic Ridge Basalt}
\label{ssec:MORB}
Measurements by probes and orbiter missions infer that rocks of basaltic composition are very common in the solar system as they can also be found on Mars, Venus, Mercury and on our Moon \citep[e.g.][]{Grotzinger1475, 2017SSRv..212.1511G, 2019RAA....19...52W}.
The major difference of these rock compositions is the \ce{Fe} content, which is related to the planetary mass, causing a different degree of differentiation \citep{2012AREPS..40..113E}.
Another factor is the accretion history of the planet as well as the interior chemistry, causing different interior redox states.
On Earth, basalt is most common in the oceanic crust.
Therefore, we used the mean composition of the Mid Oceanic Ridge Basalt (MORB) from \citet{Arevalo201070} for our analysis.

Figure~\ref{fig:Abundances} and Table~\ref{tab:abundances} show that MORB, in general, has element abundances similar to CC and BSE, but is enriched in S, Ca and Ti, and poor in N. The abundance of Mg in MORB is significantly larger than in CC, but less than in BSE. We restrict our model to the same elements as in the previous section (H, C, N, O, F, Na, Mg, Al, Si, P, S, Cl, K, Ca, Ti, Cr, Mn, Fe and their corresponding ions).

The results of this model are shown in Fig.~\ref{fig:GGchem-MORB}. Important liquids at high temperatures are found to be \ce{SiO2}[l], \ce{FeO}[l], \ce{MgSiO3}[l], \ce{Al2O3}[l], \ce{Na2SiO3}[l], \ce{MgTi2O5}[l], \ce{CaO}[l] and \ce{MgAl2O4}[l], with the first condensate being \ce{Al2O3}[l] at 5150\,K in this model. At lower temperatures, MORB shows a particularity rich solid composition in our model, including  
\ce{SiO2}[s] ({\sl quartz}), 
\ce{MgSiO3}[s] ({\sl enstatite}), 
\ce{FeTiO3}[s] ({\sl ilmenite}),
\ce{Fe2SiO4}[s] ({\sl fayalite}),
\ce{NaAlSi3O8}[s] ({\sl albite}), 
\ce{CaTiSiO5}[s] ({\sl sphene}),
\ce{CaMgSi2O6}[s] ({\sl diopside}),
\ce{Fe3O4}[s] ({\sl magnetite}),
\ce{CaAl2Si2O8}[s] ({\sl anorthite}), 
\ce{Al2SiO5}[s] ({\sl kyanite}),
\ce{FeO}[s] ({\sl ferropericlase}),
\ce{Mg2SiO4}[s] ({\sl fosterite}),
\ce{NaAlSi2O6}[s] ({\sl jadeite}),
\ce{Ca3Al2Si3O12}[s] ({\sl grossular}),
\ce{CaSiO3}[s] ({\sl wollastonite})
\ce{Fe2TiO4}[s] ({\sl ulvospinel}),
\ce{Ca2SiO4}[s] ({\sl larnite}),
\ce{FeS}[s] ({\sl troilite}) and 
\ce{C}[s] ({\sl graphite}).

The atmosphere over MORB is predicted to be more reducing than over CC and BSE, with \ce{O2} only playing a minor role.
Between temperatures of about 1300\,K and 3500\,K, the atmosphere consists mainly of \ce{H2O} and \ce{SO2} with a particle ratio of about 6:1, according to the assumed element abundances of H and S.
At higher temperatures, the gas mainly consists of \ce{SiO} molecules and O atoms.
At lower temperatures, \ce{FeS}[s] condenses at $T\!<\!1300$\,K, removing \ce{SO2} from the gas phase.
The following phyllosilicates become stable below about 750\,K:
\ce{Mg3Si4O12H2}[s] ({\sl talc}), 
\ce{FeAl2SiO7H2}[s] ({\sl Fe-chloritoid}) and 
\ce{KMg3AlSi3O12H2}[s] ({\sl phlogopite}).
These phyllosilicates remove the water from the atmosphere and lead to a narrow temperature interval within which \ce{CO2} becomes the most abundant gas species. They also prevent the formation of liquid water at lower temperatures. At $T\!<\!650$\,K, \ce{C}[s] ({\sl graphite}) condenses, leaving behind an \ce{N2}-rich atmosphere. 

\subsection{CI chondrite abundances}
\label{ssec:meteorite}
\begin{figure}[!t]
\hspace*{-4mm}
\includegraphics[width = 94mm, page=8] {./graphs/CIchond/ggchem_100.pdf}\\[-1ex]
\hspace*{-3mm}
\includegraphics[width = 92mm] {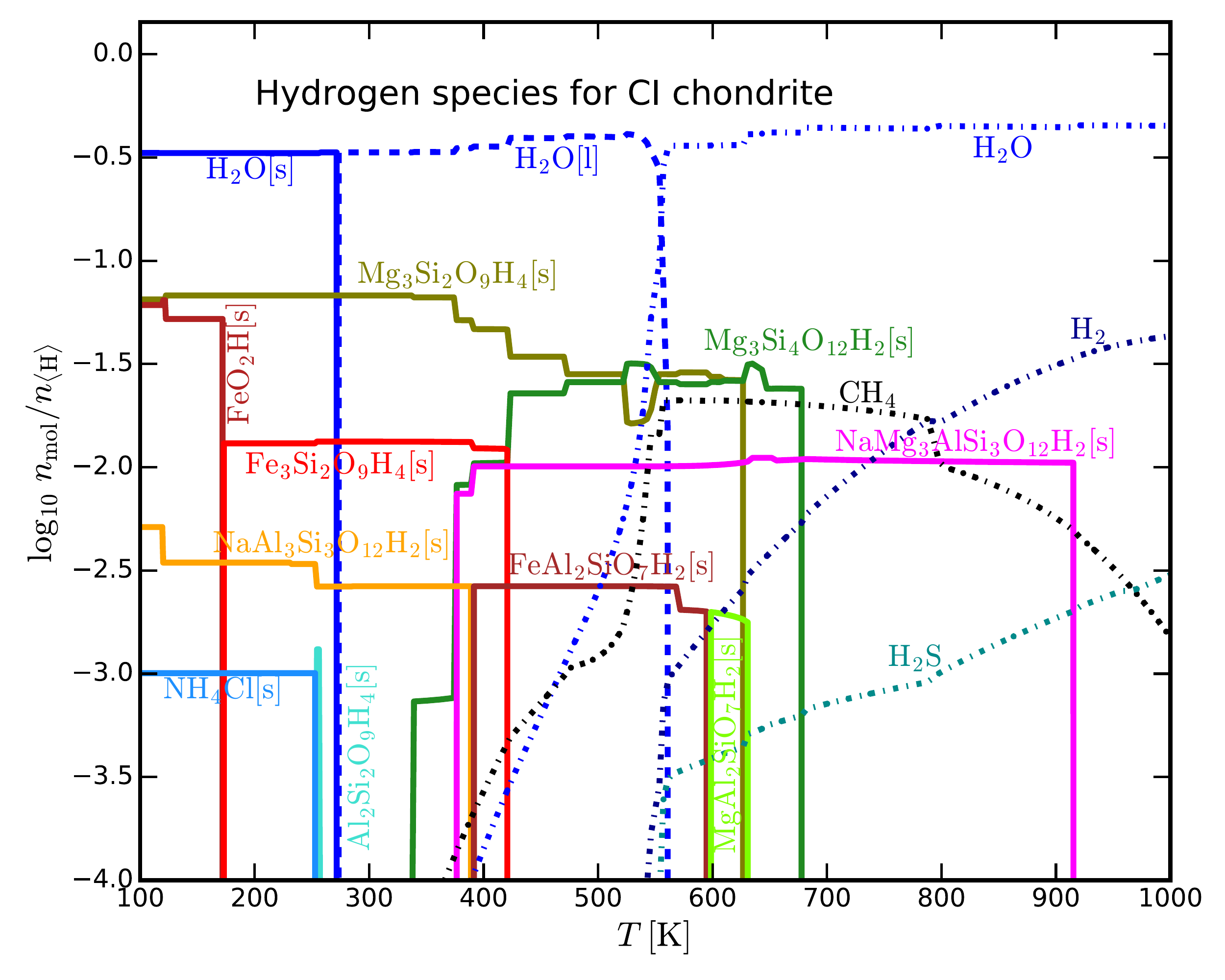}\\[-4ex]
\caption{Results for carbonaceous chondrite (CI) total element abundances at constant pressure $p\!=\!100\,$bar. Phyllosilicates are included in the model. {\bf Top panel:}  gas phase concentrations ($n_\mathrm{mol}/n_\mathrm{tot}$) between 100\,K and 5000\,K. All species with maximum log concentration $>\!-1.4$ are shown. {\bf Bottom panel:} gaseous and condensed species that contain hydrogen per H-nucleus ($n/n_{\langle H\rangle}$) between 100\,K and 1000\,K, note the different scaling.
The linestyles correspond to the different categories of condensates: silicates (solid), liquids (dotted), phyllosilicates (short dashed), other H-bearing species (long dashed), and other condensates (dash dotted).}
\label{fig:GGchem-CIcond}
\end{figure}

Chondrites are believed to be remainders of the formation period of the planets and thus can provide insights into the composition of the planets at very early stages.
It remains a matter of debate whether it is still possible today to find meteorites that resemble the building blocks of Earth, since their isotope ratios do not match those on Earth \citep[see e.g.][]{2002Natur.416...39D, 1812.11717}.
However, the isotope ratios of the planets can be explained by a mixture of different types of chondrites.

In order to investigate potential atmospheres based on these primitive remainders of the planet formation, Fig.~\ref{fig:GGchem-CIcond} shows the results for CI chondrite element abundances, based on the {\sl Orgueil} meteorite \citep[see][and Table~\ref{tab:abundances}]{2009LanB...4B..712L}.
This meteorite has an extraordinarily primitive composition, which is rich in Fe, Mg and S, but has only little Si, Ca and Al.
The volatile elements H, C and N are significantly more abundant than for CC, BSE and MORB (see Fig.~\ref{fig:Abundances}).

Due to these differences, the liquid and solid composition is very different from all other cases studied so far.
The first condensate  \ce{MgAl2O4}[l] appears at $T\!\approx\!4500\,$K at 100\,bar, while other abundant liquids are \ce{SiO2}[l], \ce{FeO}[l], \ce{FeS}[l], \ce{MgSiO3}[l], \ce{Mg2SiO4}[l] and \ce{MgO}[l].
The most important solid condensates are
\ce{SiO2}[s] ({\sl quartz}),
\ce{FeS}[s] ({\sl troilite}),
\ce{MgSiO3}[s] ({\sl enstatite}),
\ce{NaAlSi3O8}[s] ({\sl albite}),
\ce{C}[s] ({\sl graphite}),
\ce{FeS2}[s] ({\sl pyrite}),
\ce{Fe3O4}[s] ({\sl magnetite}),
\ce{FeO}[s] ({\sl ferropericlase}),
\ce{Fe2SiO4}[s] ({\sl fayalite}) and 
\ce{NaAlSiO4}[s] ({\sl nepheline}).
A number of carbonates becomes stable below about 500\,K, among them
\ce{FeCO3}[s] ({\sl siderite}),
\ce{MgCO3}[s] ({\sl magnesite}) and
\ce{MnCO3}[s] ({\sl rhodochrosite}).
At temperatures below 400\,K to 900\,K, hydration is very common, leading to the formation as phyllosilicates as
\ce{Mg3Si4O12H2}[s] ({\sl talc})
\ce{NaMg3AlSi3O12H2}[s] ({\sl sodaphlogopite}),
\ce{Mg3Si2O9H4}[s] ({\sl lizardite}) and
\ce{Fe3Si2O9H4}[s] ({\sl greenalite}).
Despite the formation of those phyllosilicates, there is still enough hydrogen available for liquid and solid water to become thermally stable in this model, \ce{H2O}[l] at 560\,K and \ce{H2O}[s] at 271\,K.  Interestingly, \ce{FeO2H}[s] ({\sl goethite}) becomes stable at about 170\,K.

The gas phase mainly consists of \ce{H2O} for almost all temperatures considered, followed by \ce{CO2}, \ce{CO} and \ce{H2}, and the sulphur molecules \ce{H2S} and \ce{SO2}.
At very high temperatures, $T\!>\!4500\,$K, \ce{O}, \ce{H}, \ce{OH} and \ce{SiO} and \ce{CO} become more abundant.
Once the phyllosilicates, graphite and the carbonates have formed around 500\,K, the gas phase is dominated by \ce{N2}.
Although \ce{NH4Cl} ({\sl ammonium chloride}) becomes stable around 250\,K in this model, the chlorine abundance is not large enough here to exhaustively consume \ce{N2}.

In the bottom panel of Fig.~\ref{fig:GGchem-CIcond} we show the hydrogen bearing species between 100\,K and 1000\,K.
The different phases of \ce{H2O} incorporate most of the H atoms.
The first H bearing condensate is 
\ce{NaMg3AlSi3O12H2}[s] {\sl (sodaphlogopite)} at about 910\,K, followed by
\ce{Mg3Si4O12H2}[s] {\sl (talc)}, 
\ce{MgAl2SiO7H2}[s] {\sl (Mg-chloritoid)}, 
\ce{Mg3Si2O9H4}[s] {\sl (lizardite)} and 
\ce{FeAl2SiO7H2}[s] {\sl (Fe-chloritoid)},
before \ce{H2O}[l] condenses at about 550\,K.
The amount of H kept in phyllosilicates increases steadily to lower temperatures, but is insufficient to suppress the thermal stability of liquid water. 

The CI chondrite composition is the only rocky element composition considered in this paper that produces liquid and solid water without the need for additional hydrogen and oxygen.
Although other rock compositions produce large amounts of gaseous water as well, phyllosilicates are thermodynamically more favourable than the liquid water, and hence inhibits the stability of liquid water for those other element abundances in phase equilibrium. However, the CI chondrites are hydrated so much that even after the formation of the phyllosilicates, there is still some water left to condense.

\subsection{Polluted white dwarf abundances}
\label{ssec:PWD}
\begin{figure}[!t]
\hspace*{-4mm}
\includegraphics[width = 94mm, page=8] 
  {./graphs/PWD/ggchem_100.pdf}\\[-1ex]
\hspace*{-3mm}
\includegraphics[width = 94mm] 
  {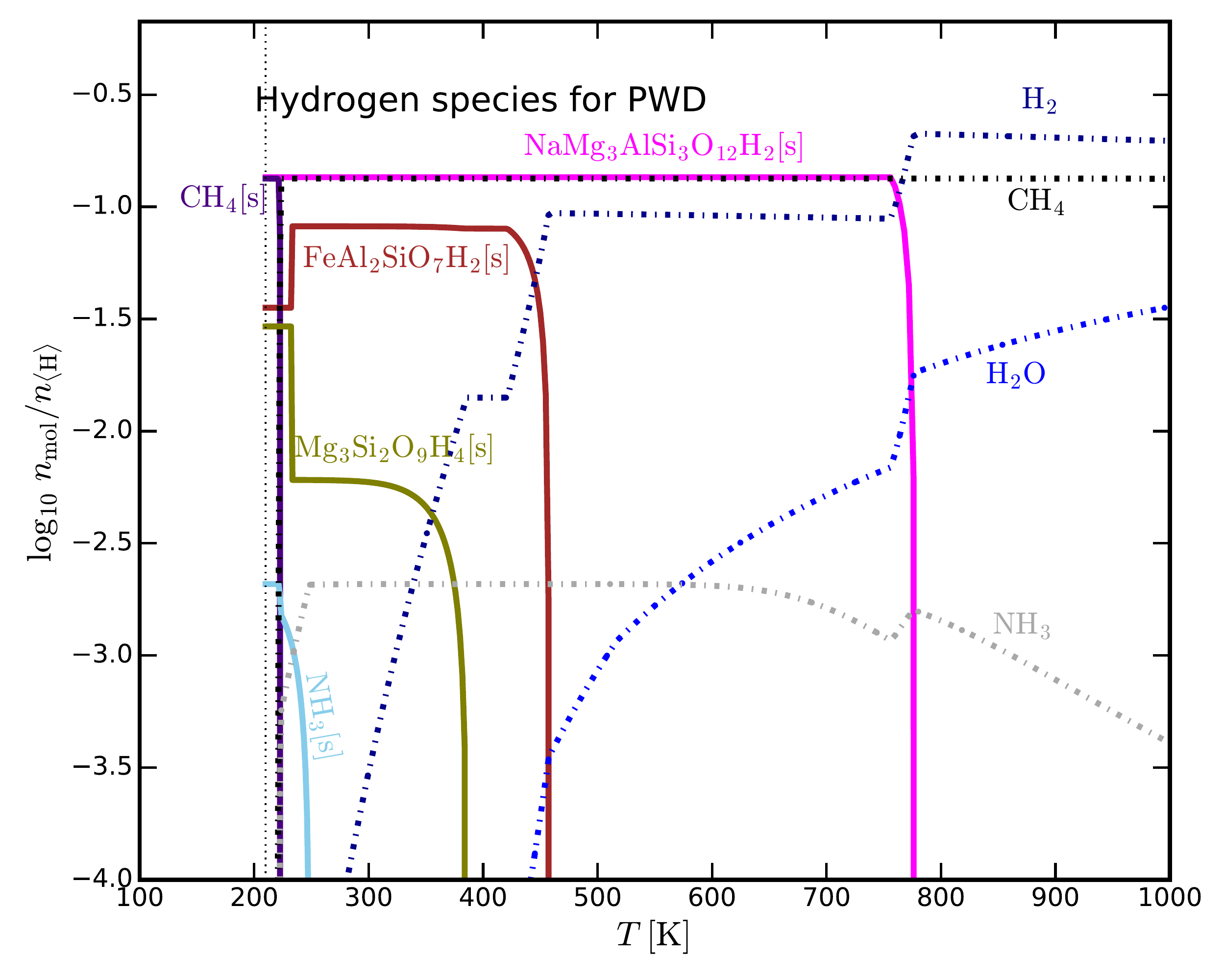}\\[-4ex]
\caption{Results for {\sl Polluted White Dwarf (PWD)} total element abundances at constant pressure $p\!=\!100$\,bar. Phyllosilicates are included as possible condensates.
{\bf Top panel:} gas phase concentrations ($n_\mathrm{mol}/n_\mathrm{tot}$) between 100\,K and 5000\,K. All species with maximum $\log n_\mathrm{mol}/n_\mathrm{tot}\! >\!-1.4$ are shown. {\bf Bottom panel:} gaseous and condensed species that contain hydrogen are plotted with their concentration per H-nucleus $n/n_{\langle H\rangle}$ between 100\,K and 1000\,K (note the different scalings).
}
\label{fig:GGchem-PWD}
\end{figure}

The bulk composition of an exoplanet can be studied in terms of its mean density, which is determined by the planet's mass and diameter. These quantities can be measured by a combination of radial velocity and transit observations \citep[e.g.][]{2014ApJ...783L...6W, 2019MNRAS.484.3731R}.
A new, powerful and potentially more direct approach to measure the bulk composition of exoplanets is to determine the element composition in the atmospheres of polluted white dwarfs.

White dwarfs (WDs), the burnt-out cores of low-mass stars which become visible only after the terminal ejection of a planetary nebulae, usually have pure hydrogen/helium atmospheres, because all elements heavier then He settle down quickly\linebreak ($\sim$\,$10^5\,$yrs) in the extremely strong gravitational field of the object \citep[see e.g.][]{Paquette1986,Koester2009}.
Nevertheless, some WDs show an enrichment in elements like O, Mg, Al, Si, Ca, Fe, Si, and C \citep[e.g.][]{2016MNRAS.459.3282W,2016MNRAS.463.3186F}.
These white dwarfs with metal absorption line features are hence called {\it Polluted White Dwarfes} (PWDs).
The observed relative metal abundances in these PWDs are comparable to the Earth's composition with some deviations, see references in \citet{2019MNRAS.487..133W} and Fig.~\ref{fig:Abundances}.
The most plausible reason for such enrichment with heavy elements in a white dwarf's atmosphere is the accretion of planetesimals or planets from the white dwarf's planetary system.
The challenge is, however, that the datasets from PWDs are lacking some important elements as their detection is presently very challenging.

In Table~\ref{tab:abundances} we have listed the measurements for the WD GD~362 by \citet{2007ApJ...671..872Z}, which shows a composition that is particularly Fe-rich, see Fig.~\ref{fig:Abundances}. 
Unfortunately, the abundances of \ce{H} and \ce{He} as pollutant are impossible to derive this way, since these elements originate from the WD atmosphere itself.
Thus, we excluded \ce{He} from our calculations.
Furthermore the values for \ce{C}, \ce{N} and \ce{O} are only provided as upper limits.
In our model, we have therefore used averaged abundances for elements H, C, N and O, computed from mean values of the logarithmic abundances from the columns denoted by 'CC Schaefer', 'BSE Schaefer', 'MORB' and 'CI meteorite' in Table~\ref{tab:abundances}.
This results in $\epsilon_{\rm H}=6.218$, $\epsilon_{\rm C}=5.344$, $\epsilon_{\rm N}=3.5365$ and $\epsilon_{\rm O}=7.903$.
In this work, we only investigate element abundances from a single PWD which has inferred abundances for a large number of elements allowing a more diverse atmosphere and condensate composition.
An in-depth analysis of multiple elemental abundances inferred from PWD is beyond the scope of this paper.
The origin of the pollutant material range from complete planets, stripped planetary cores, comets to gas giants \citep[see e.g.][]{2018MNRAS.479.3814H, 2020arXiv200104499B}.

We have included 17 elements in our model (H, C, N, O, Na, Mg, Al, Si, Ca, Ti, V, Cr, Mn, Fe, Ni, Cu).
Some important elements have been disregarded, because their element abundances are not available, in particular F, P, S, Cl and K.
We also excluded \ce{Sc} and \ce{Co} because the condensate data in our model is not reliable for these elements.
{\sc GGchem} finds 194 gas species and 142 condensed species in its database, 23 of them being liquids, for the included 17 elements.

The results of our phase-equilibrium model are shown in Fig.~\ref{fig:GGchem-PWD}.
At $T\!\ga\!3700\,$K, \ce{Fe} is the most abundant gas species, followed by \ce{SiO} and \ce{O}.
Additionally, \ce{O2}, \ce{FeO}, \ce{Mg} and \ce{MgO} are abundant in the gas, but oxygen never becomes the dominant gas species as it is kept in large quantities in \ce{FeO}[l].

At $T$\,$\sim$\,$4900\,$K, \ce{CaO}[l] and \ce{MgAl2O4}[l] are the first condensates.
Other relevant liquids are \ce{FeO}[l], \ce{MgO}[l], \ce{MgSiO3}[l], \ce{MgTiO3}[l] and \ce{Fe}[l], before \ce{CaO}[l] solidifies as \ce{Ca2SiO4}[s] at $T$\,$\sim$\,$3900\,$K.
Further condensates are \ce{Ni}[l] and \ce{Ti4O7}[s], before the condensation of \ce{SiO2}[l], \ce{CaSiO3}[s], \ce{Ca2SiO4}[s], \ce{Mn2SiO4}[s], \ce{Na2SiO3}[l] and \ce{Ca2MgSi2O7}[s] in the temperature range $2900\,{\rm K}\!<\!T\!<\!3400\,$K causes the atmospheric composition to change from an \ce{SiO} dominated atmosphere to an \ce{H2} dominated atmosphere with \ce{H2O} and \ce{CO} being further abundant gas species.

At $T$\,$\la$\,$1300\,$K, \ce{CO} transforms into \ce{CH4}, which causes the \ce{H2O} concentration to decrease, which leaves an atmosphere rich in \ce{H2}, \ce{CH4} and \ce{H2O}.
The thermal stability of the phyllosilicates 
\ce{NaMg3AlSi3O12H2}[s] {\sl (sodaphlogopite)} at ${T\!\la\!770\,\mathrm{K}}$, \ce{FeAl2SiO7H2}[s] {\sl(iron-chloritoid)} at ${T\!\la\!460\,\mathrm{K}}$ and 
\ce{Mg3Si2O9H4}[s] {\sl(lizardite)} at ${T\!\la\!380\,\mathrm{K}}$ removes \ce{H2} and \ce{H2O} from the gas phase, eventually leading to a pure \ce{CH4} atmosphere.
The trace gas \ce{NH3} condenses at ${T\!\la\!250\,\mathrm{K}}$.
The last gas species to become stable as a condensate is \ce{CH4} at ${T\!\la\!220\,\mathrm{K}}$.
For even lower temperatures, no species remain stable in the gas, and therefore no physical solution is possible for a $p\!=\!100$\,bar atmosphere.
The formation of phyllosilicates inhibits once again the formation of water as a condensate.

The PWD element abundances are particularly rich in \ce{Fe}, 
which binds oxygen from the atmosphere to form additional 
Fe-bearing solid compounds.
Hence, increasing the total Fe abundance causes a more reducing atmosphere.
The same effect can be achieved by increasing the Fe content in other 
element mixtures.
For example, considering BSE abundances, an increase for the Fe abundance by 1 wt\,\% causes the atmosphere to change from a mixture of \ce{H2O}, \ce{CO2} and \ce{SO2} to a mixture of \ce{H2}, \ce{H2O} and \ce{CH4} (see Fig.~\ref{fig:GGchem-BSE+Fe}).
The results for \ce{Fe}-enriched BSE abundances and PWD abundances are indeed very similar.
Thus, a reducing atmosphere is expected for planets with increased total iron abundance in the crust, or could be caused by the late delivery of iron-rich bodies to the planet surface.

\begin{figure*}[!th]
\centering
\includegraphics[width = .49\linewidth]{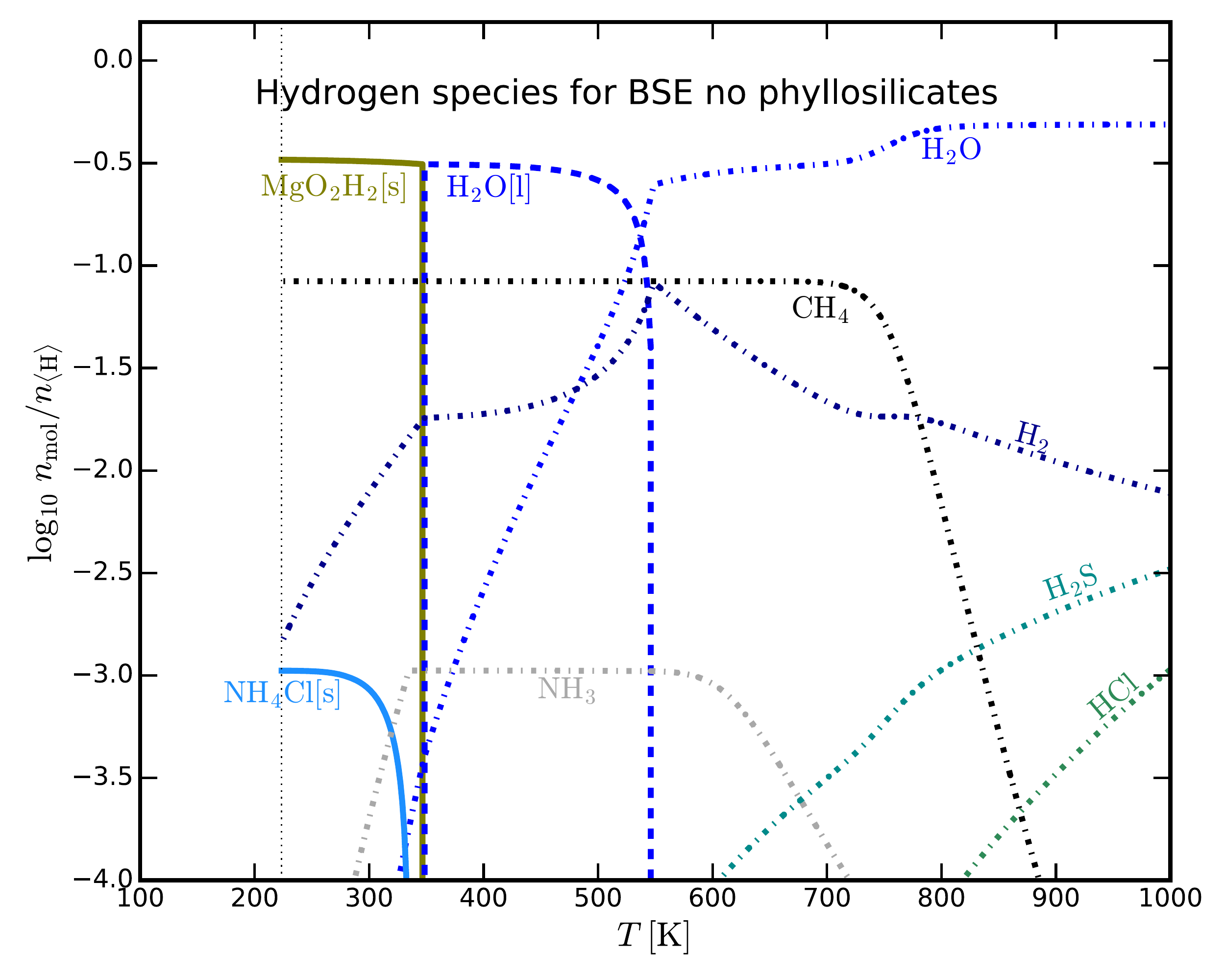}
\includegraphics[width = .49\linewidth]{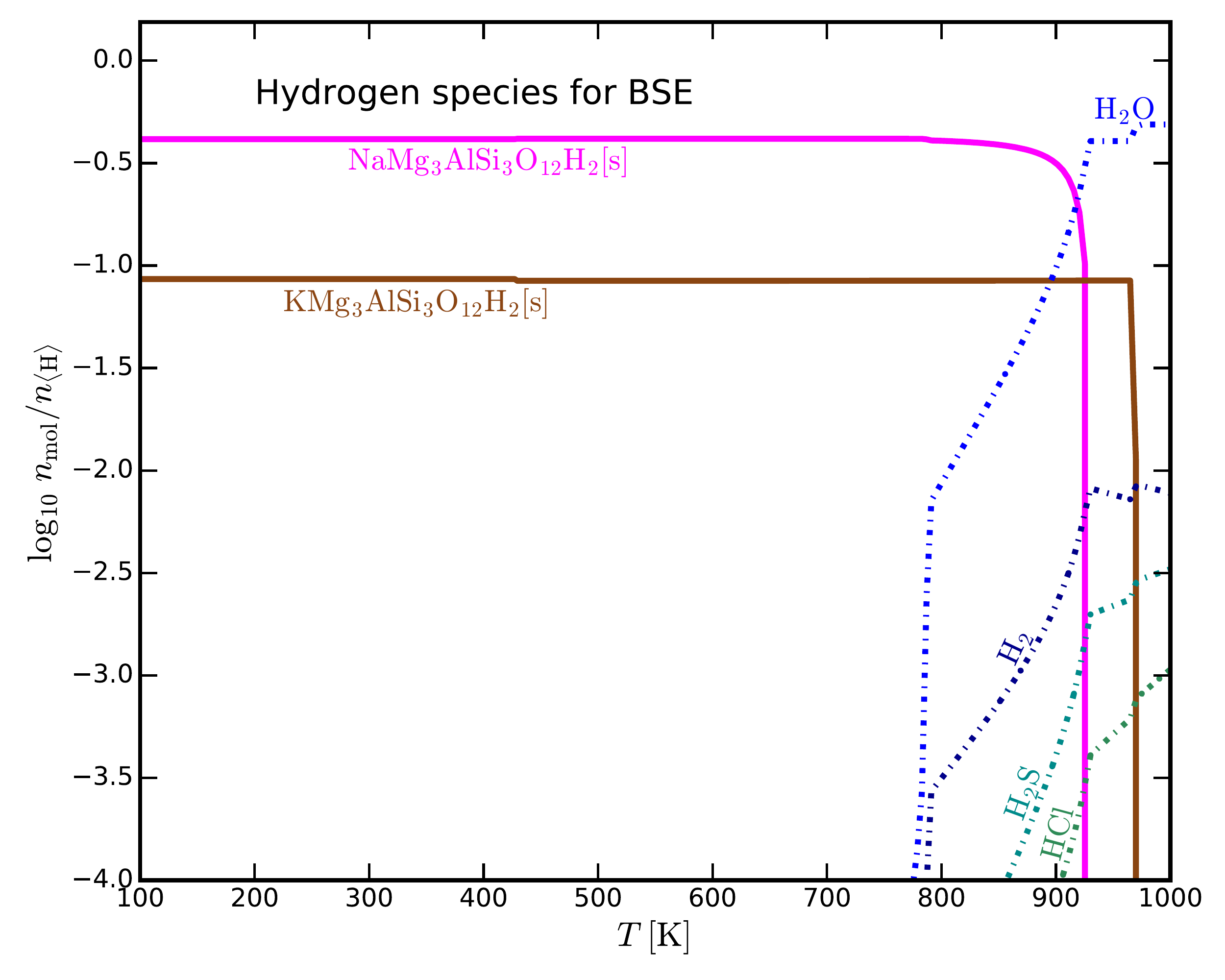}\\
\includegraphics[width = .49\linewidth]{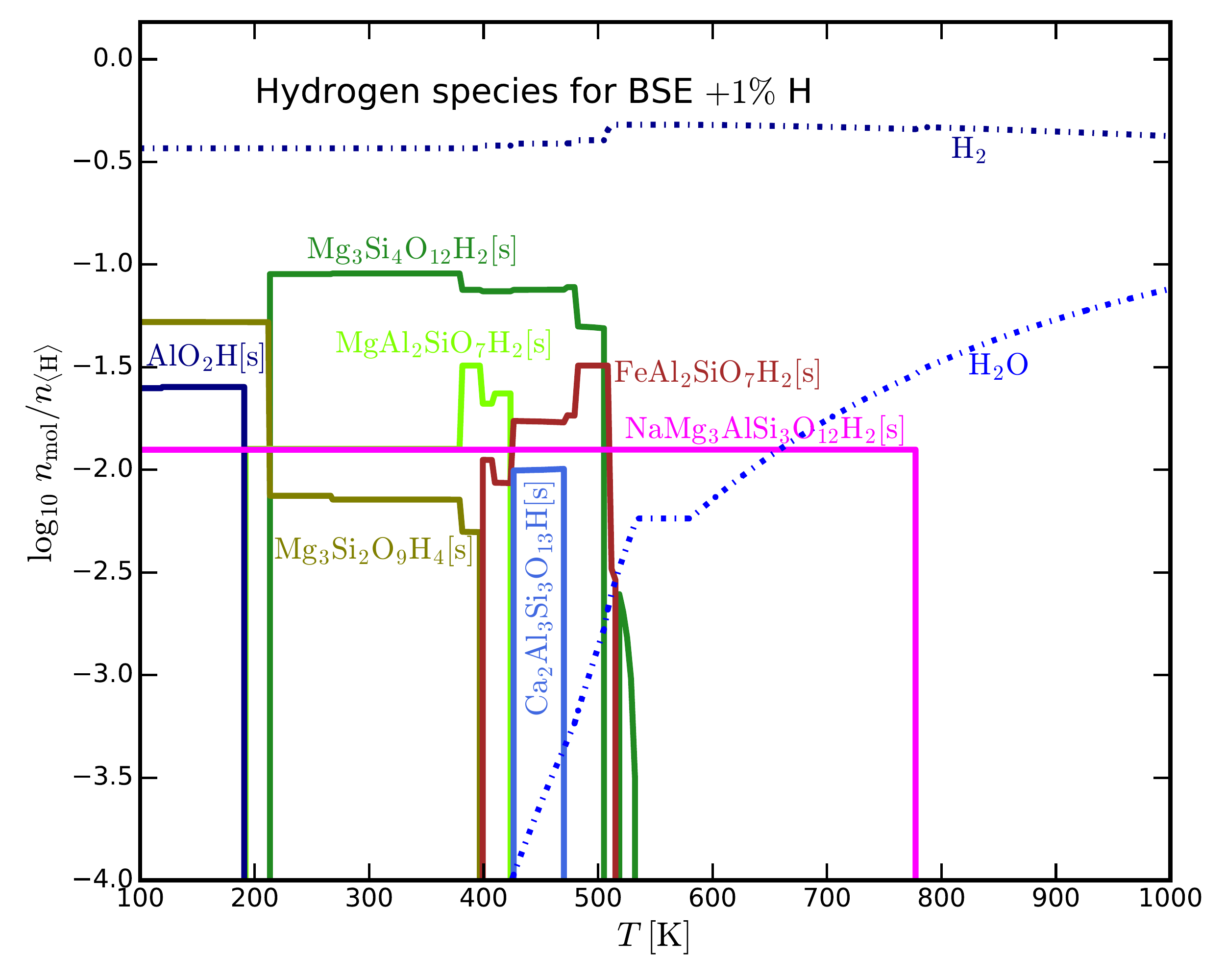}
\includegraphics[width = .49\linewidth]{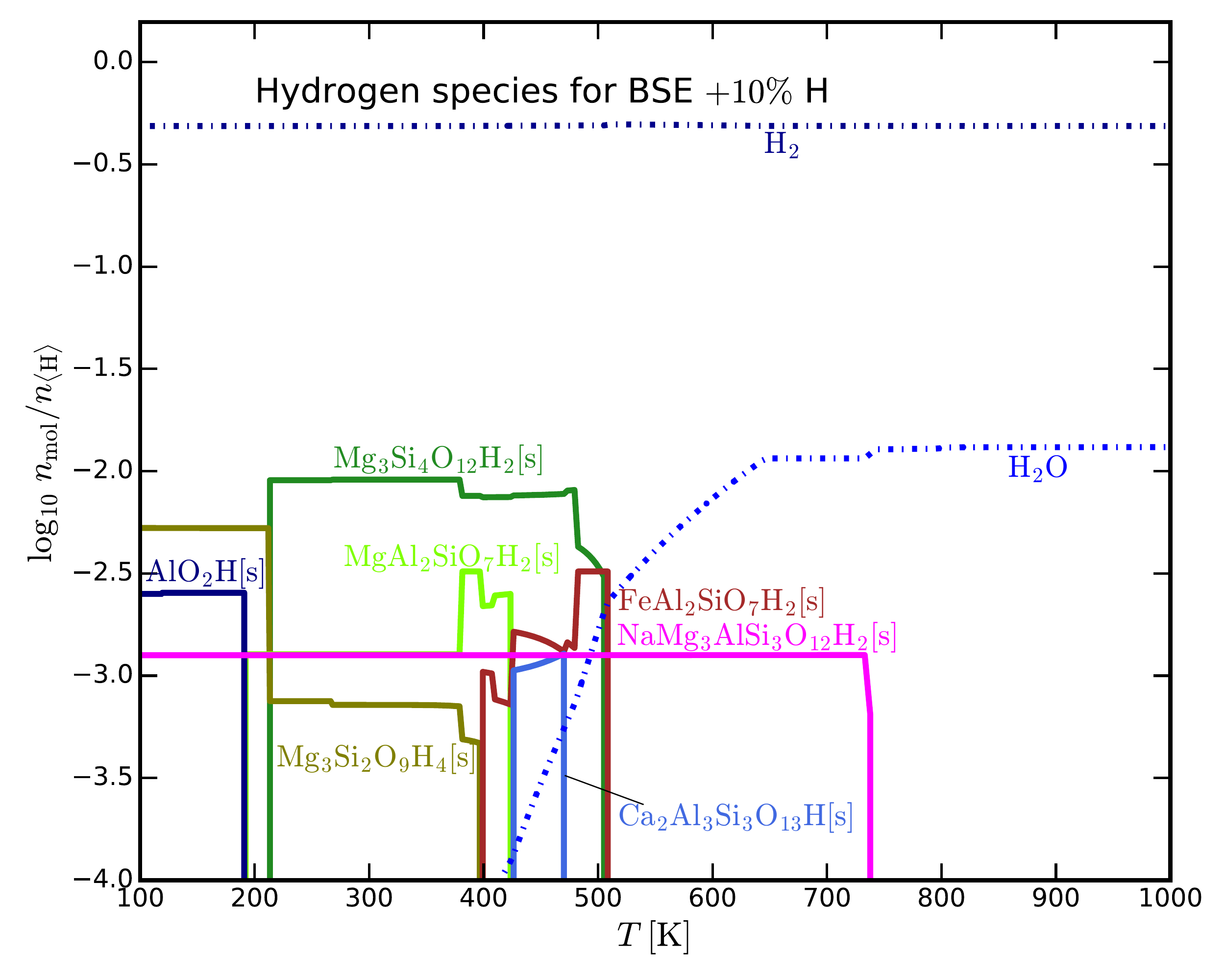}\\
\includegraphics[width = .49\linewidth]{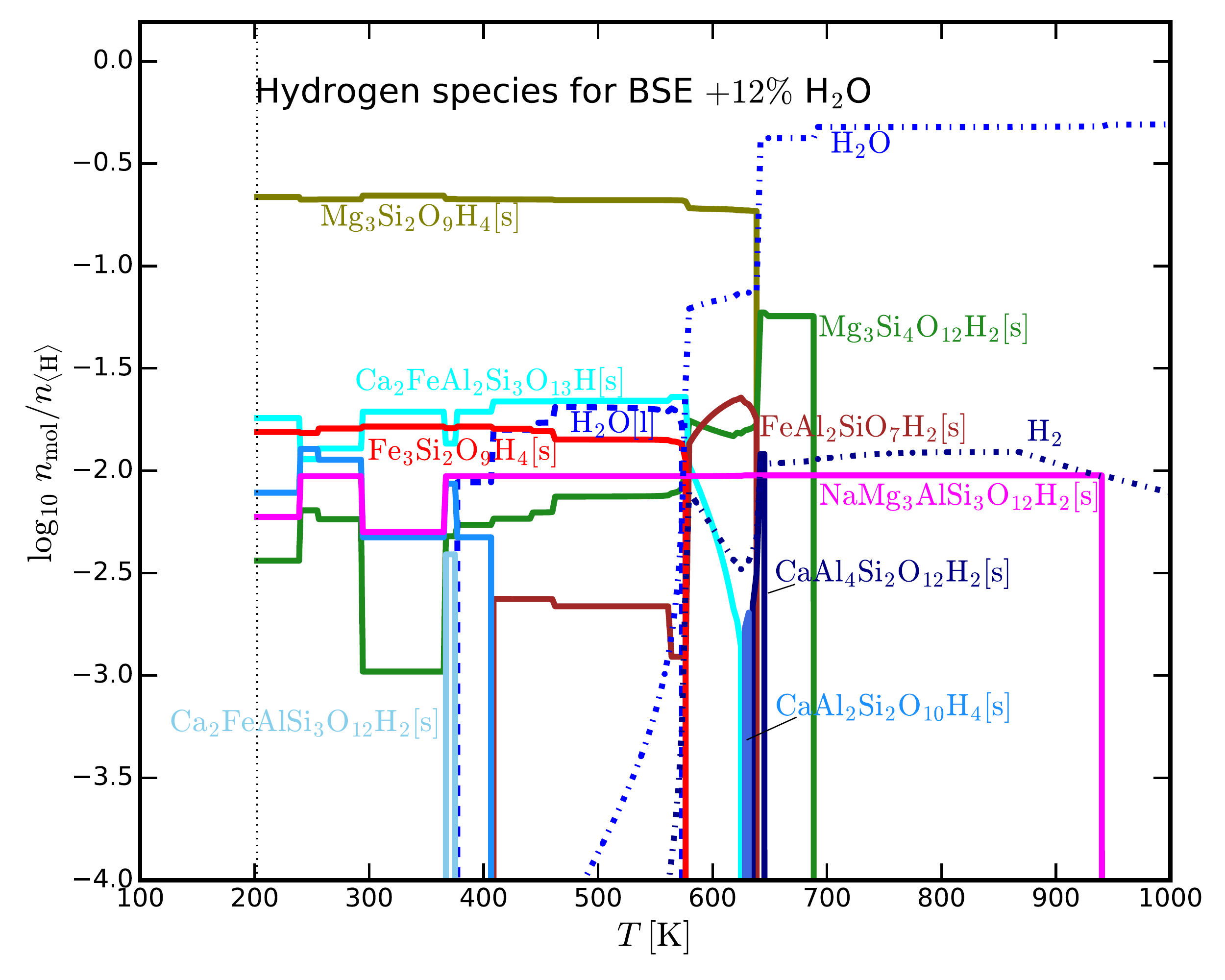}
\includegraphics[width = .49\linewidth]{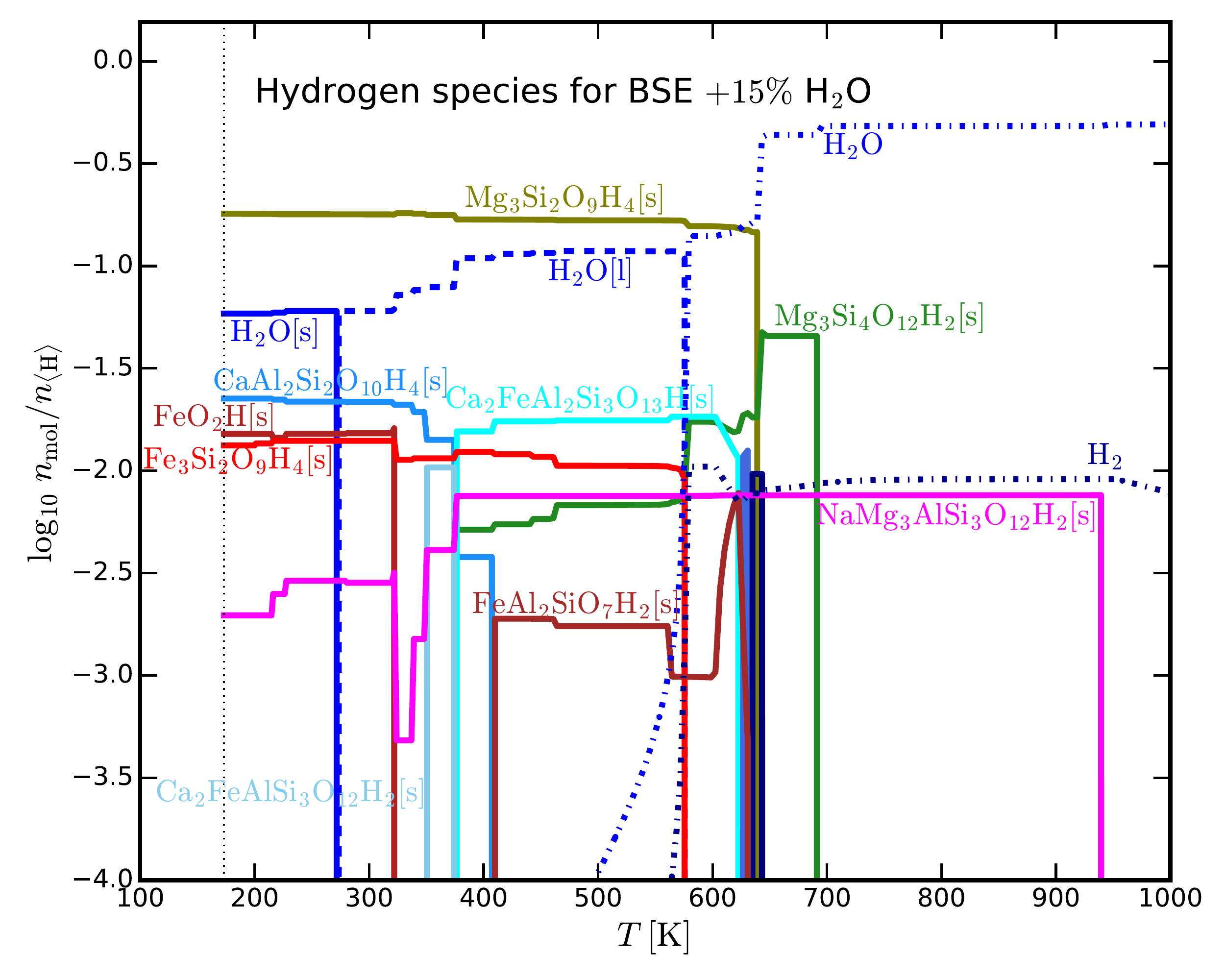}
\caption{Distribution of hydrogen among the different gas and condensed species. All models are run at 100\,bar and for the following abundances:
{\bf Top left panel:} BSE without phyllosilicates;
{\bf Top right panel:} BSE;
{\bf middle left panel:} BSE + 1.339 wt\,\% H;
{\bf middle right panel:} BSE + 15 wt\,\% H;
{\bf bottom left panel:} BSE + 12 wt\,\% \ce{H2O};
{\bf bottom right panel:} BSE + 15 wt\,\% \ce{H2O}.
The threshold concentration for species to be shown is set to $10^{-3}$ for all abundances except for BSE + 10\% H ($10^{-3.5}$).
}
\label{fig:BSE_water}
\vspace*{8mm}
{\ }
\end{figure*}

\begin{figure}[!hbtp]
\hspace*{-3mm}
\includegraphics[width=90mm,trim=0 18 18 0,clip] {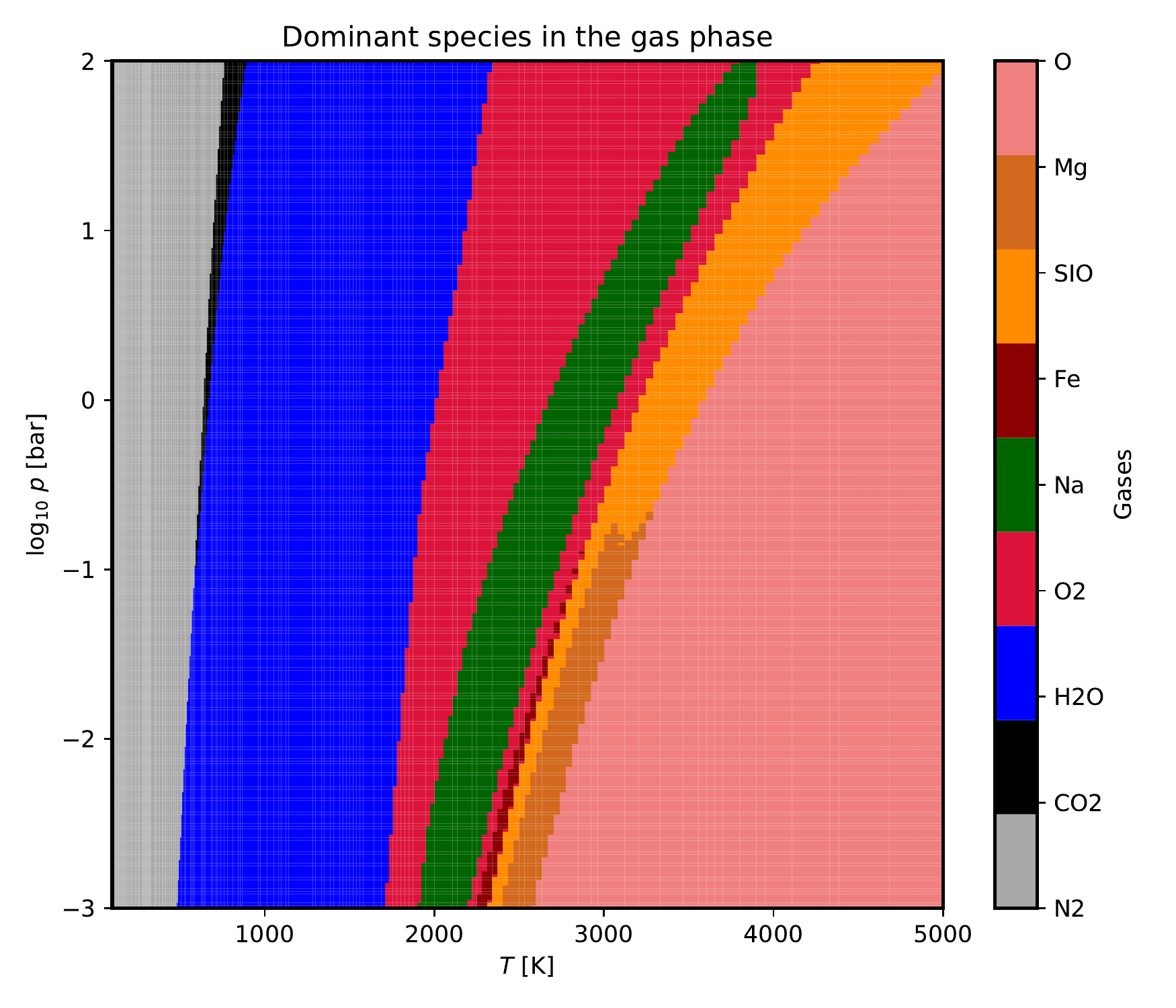}\\[-1ex]
\hspace*{-4mm}
\includegraphics[width=93mm, page=8] {./graphs/BSE_pressure/ggchem_1.pdf}\\[-1ex]
\hspace*{-4mm}
\includegraphics[width=93mm, page=8] {./graphs/BSE_pressure/ggchem_0001.pdf}\\[-4ex]
\caption{Results for BSE total abundances at different pressures.
{\bf Upper panel:} dominant gas species with respect to the highest concentration in the $p-T$ plane.
{\bf Middle panel:} gas concentrations ($n_\mathrm{mol}/n_\mathrm{tot}$) at $p\!=\!1$\,bar.
{\bf Lower panel:} gas concentrations ($n_\mathrm{mol}/n_\mathrm{tot}$) at $p\!=\!0.001$\,bar.
The transitions in composition are caused by phase changes as 
listed in Tab.~\ref{tab:condensates}.
Further pressure levels are shown in Fig.~\ref{fig:p-T_plane_BSE_2}.}
\label{fig:p-T_plane_BSE}
\vspace*{-10mm}
\end{figure}

\section{Stability of condensed water}\label{sec:abundance_var}
The occurrence of surface water is believed to be one of the necessary conditions for the emergence of life as we know it \citep[e.g.][]{Westall2018}.
Therefore, we investigate the potential stability of water in an equilibrium crust-atmosphere model.

The results of our models, as discussed in the previous sections, show that the formation of phyllosilicates can inhibit the formation of liquid and solid water for most rock compositions, except for the CI chondrite abundances.
This raises the question by how much we need to change the total element abundances in order to allow for water to condense.
We therefore have carried out additional simulations with altered element abundances based on the Bulk Silicate Earth (BSE) dataset.

Crucial for the water formation is the presence of hydrogen.
We have therefore tested two approaches:
(i) increasing the \ce{H} abundance, and
(ii) increasing the \ce{H} and \ce{O} abundances with particle ratio 2:1.
The respective elements are added to the BSE abundances before normalising them to 1.
Fig.~\ref{fig:BSE_water} shows our results for temperatures between 100\,K and 1000\,K for these models.

The two top panels once again demonstrate the effect of the phyllosilicates. 
Only if the phyllosilicates are artificially removed from our list of condensates, we find liquid water to be thermally stable between about 350\,K and 550\,K.
If the phyllosilicates are included, however, in particular \ce{NaMg3AlSi3O12H2}[s] ({\sl sodaphlogopite}) and \ce{KMg3AlSi3O12H2}[s] ({\sl phlogopite}), hydrogen is very efficiently removed from the gas phase, and liquid water cannot become thermally stable (upper right panel of Fig.~\ref{fig:BSE_water}).

With additional \ce{H} only, we have been unable to find models where liquid water would be thermally stable.
We depict the concentrations of the hydrogen species in the middle panels of Fig.~\ref{fig:BSE_water} after increasing the hydrogen abundance by 1\% and 10\%, respectively.
The dominant \ce{H} bearing species in both cases is found to be \ce{H2}.
In comparison to the unaltered BSE model, however, a larger variety of phyllosilicates is found to form: 
\ce{NaMg3AlSi3O12H2}[s] ({\sl sodaphlogopite}), 
\ce{FeAl2SiO7H2}[s] {\sl(iron-chloritoid)}, 
\ce{Ca2Al3Si3O13H}[s] {\sl(clinozoisite)}, 
\ce{Mg3Si4O12H2}[s] {\sl (talc)}, 
\ce{MgAl2SiO7H2}[s] {\sl (Mg-chloritoid)}, 
\ce{Mg3Si2O9H4}[s] {\sl(lizardite)} and 
\ce{AlO2H} {\sl(diaspore)}.

With additional \ce{H} and \ce{O}, the formation of liquid and solid water
succeeds in our models, see lower panels of Fig.~\ref{fig:BSE_water}. 
Since the mass ratio O:H is 16:1, the inclusion of 1 wt\,\% water results in an increase of 1/9 wt\,\% for \ce{H} and 8/9 wt\,\% for \ce{O}.
We found that liquid water becomes stable after adding about 12 wt\,\% of \ce{H2O} to the total BSE element abundances, whereas 15 wt\,\% are required for the stability of solid water.
The diversity of phyllosilicates increases again.
In addition we find 
\ce{Ca2FeAl2Si3O13H}[s] ({\sl epidote}), 
\ce{CaAl2Si2O10H4}[s] ({\sl lawsonite}), 
\ce{Ca2FeAlSi3O12H2}[s] ({\sl ferri-prehnite}), 
\ce{CaAl4Si2O12H2}[s] ({\sl margarite}) and 
\ce{Fe3Si2O9H4}[s] ({\sl greenalite}).
For $T\!<\!200\,$K, no atmosphere is found to be stable, since all included elements are thermally stable in condensates.

This analysis shows that phyllosilicates can act as a reservoir for capturing a certain amount of water.
By adding sufficient hydrogen {\sl and} oxygen, it is possible to saturate the phyllosilicates and to have liquid and solid water as stable condensates.
The added water in our model could be an indicative of an additional delivery of water to the planet, for example via the incorporation of icy comets. However, there are other possible explanations.
For example, the existence of phyllosilicates is thermodynamically impossible in the hot core and the overwhelming part of the mantle.
The large amount of water that had once been present in this matter is likely to have been driven out into the crust and atmosphere during planet evolution.
This way, there is plenty of water available to saturate the phyllosilicates in the crust {\sl and} to have excess water to form an ocean.

\section{Pressure variation}
\label{sec:pressure}

In the previous sections only a particular, fixed pressure of 100\,bar was considered.
However, the atmospheric pressures at the bottom of rocky planets can vary by orders magnitude, from several mbars to about 100\,bars, comparable to our Solar System planets Mars and Venus, respectively.
The hot atmospheres of very young rocky exoplanets might even have atmospheric pressures of multiple 100\,bars \citep[e.g.][]{DHALIWAL2018249, OLSON2018418}, following the general trend of increased vapour pressures for higher temperatures.
Therefore, we investigate the influence of the atmospheric pressure on the atmospheric and crust composition in this section.
This analysis is based on the BSE total element abundances.

In Fig.\,\ref{fig:p-T_plane_BSE} we show the most abundant gas species in the $P\!-\!T$ plane between 0.001\,bar and 100\,bar, and between 100\,K and 5000\,K.
The middle and lower panels show the atmospheric compositions at selected, constant pressures of 1\,bar and at 0.001\,bar, respectively. The plot for $p\!=\!100\,$bar was already shown in Fig.\,\ref{fig:comparison_BSE}, lower panel.
Additional pressure levels (10\,bar and 0.1\,bar) are available in the Appendix, see Fig.\,\ref{fig:p-T_plane_BSE_2}.

Generally speaking, the phase changes occurring in the atmosphere are the same, but shift to lower temperatures for lower pressures.
In particular, the temperature window in which \ce{O2} is the dominant gas species becomes narrower for lower pressures.
At 100\,bar this range is roughly 2000\,K to 4000\,K, whereas at 0.001\,bar it narrows down to about 1600\,K to 1900\,K.
The temperatures where the various condensates appear and disappear are shifted likewise, see Table\,\ref{tab:condensates}.

\begin{figure}[!t]
\hspace*{-3mm}
\includegraphics[width=94mm] {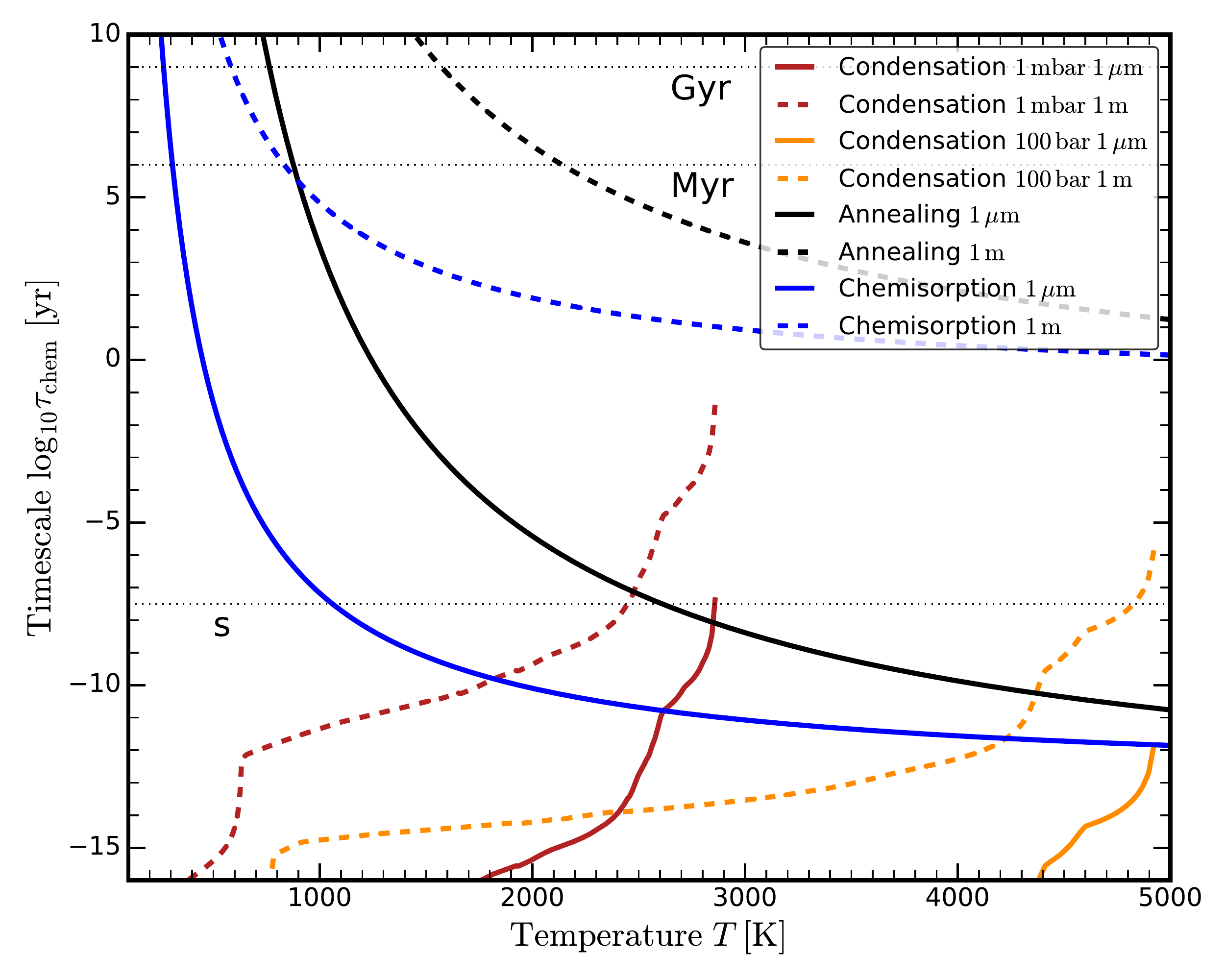}\\[-1ex]
\caption{Timescale estimates for the BSE total element abundances. The condensation timescale is shown for different pressures and length scales of $1\,\mu$m and 1\,m.
The timescales for annealing and chemisorption are shown for scales of $1\,\mu$m and 1\,m each.
The horizontal dotted lines illustrate times of 1\,s, 1\,Myr and 1\,Gyr, respectively.}
\label{fig:timescales_BSE}
\end{figure}
\section{Timescales}
\label{sec:timescales}
All calculations presented assumed that our systems have reached chemical  equilibrium.
We address this assumption by investigating three timescales.
First, the timescale for the condensation from the gas phase to form the different condensates.
Second, an annealing timescale that rearranges the condensate phase towards the thermodynamically favoured phase.
Last, we investigate the chemisorption timescale, describing the formation of phyllosilicates. 

We consider the relaxation of a state $y(t)$ towards equilibrium $y_0$ via the first-order ordinary differential equation $\mathrm{d}y/\mathrm{d}t = f(y)$, where $f$ is the 
given time derivative.
The relaxation timescale is found by considering small deviations from the equilibrium state as 
\begin{align}
    \frac{\mathrm{d}y}{\mathrm{d}t} \approx f(y_0) + \frac{\mathrm{d}f}{\mathrm{d}y} (y-y_0),
\end{align}
where $f(y_0)=0$.
The relaxation timescale is $(y-y_0)/(\mathrm{d}y/\mathrm{d}t)$, hence
\begin{align}
    \tau_\mathrm{relax} = \left(\left.\frac{\partial f}{\partial y}\right|_{y_0}\right)^{-1}. \label{eq:t_relax}
\end{align}
Close to the equilibrium state, the system relaxes as
\begin{align}
    y(t) = y_0 + (y(0)-y_0) \exp(-t/\tau_\mathrm{relax}).
\end{align}
For the gas-condensed phase transition we have $y=V_s$ and
\begin{align}
    f(V_s) &= (n - n_0) v_\mathrm{th} A_\mathrm{tot} V_1 \alpha\\
    &= (V_{s0} -V_s) v_\mathrm{th} A_\mathrm{tot} \alpha. \label{eq:f(V)}
\end{align}
where $V_s$ is the volume of a condensate $s$ per cm$^3$, $V_{s0}$ its value in phase equilibrium, $n$ the number of key molecules in the gas phase forming the condensate, and $n_0$ its value in phase equilibrium. $n$ and $V$ are connected as 
\begin{align}
    V_1 (n-n_0) = - (V_s-V_{s0}),
\end{align} 
i.e. an excess of key molecules in the gas phase means that the condensation is momentarily not complete. 
$V_1$ is the monomer volume.
$A_\mathrm{tot}$ is the total surface of all condensates and $\alpha$ the sticking coefficient, assumed to be unity.
$v_\mathrm{th}$ is the thermal velocity calculated by
\begin{align}
 v_\mathrm{th} = \sqrt{\frac{8 k_\mathrm{B} T_\mathrm{g}}{\pi m}}.
\end{align}
with the mass $m$ of the gas species containing most of the limiting element. 
The limiting element is the least abundant element in the gas phase that is included in the considered condensate.
When computing Eq.~\ref{eq:tcond}, we consider the longest timescale of all active condensates.
The identifications of the slowest condensate, its limiting element and key species have little effect on the results as they only enter via the thermal velocity which requires a species mass.
$k_\mathrm{B}$ is the Boltzmann constant and $T_\mathrm{g}$ the gas temperature.
Inserting Eq.~\ref{eq:f(V)} in Eq.~\ref{eq:t_relax} the condensation timescale $\tau_\mathrm{cond}$ follows as
\begin{align}
\tau_\mathrm{cond} = \left( v_\mathrm{th} A_\mathrm{tot} \alpha \right)^{-1} = \left( v_\mathrm{th} \frac{A_\mathrm{tot}}{V_\mathrm{tot}} V_\mathrm{tot} \alpha \right)^{-1}.\label{eq:tcond}
\end{align}
$V_\mathrm{tot}$ is the total volume of all condensates.
Equation \ref{eq:tcond} expresses the time it takes for the key molecule
of the slowest condensate to collide with an existing surface.
We approximate the surface-volume fraction $A_\mathrm{tot}/V_\mathrm{tot}$ by the dimension ratio
\begin{align}
\frac{A_\mathrm{tot}}{V_\mathrm{tot}} \propto \frac{a^2}{a^3} = \frac{1}{a},
\end{align}
with a length scale of $a$.
The inclusion of different scales will introduce factors of the order of $10^0$.
E.g. spherical symmetric grains with a radius of $a$ will have a surface-volume ratio of
\begin{align}
\frac{A_\mathrm{tot}}{V_\mathrm{tot}} = \frac{4 \pi a^2}{\frac{4}{3}\pi a^3} = \frac{3}{a}.
\end{align}
In Fig.~\ref{fig:timescales_BSE}, the condensation timescale for BSE abundances is shown for pressures of 100\,bar and 1\,mbar as well as for different length scales of $1\,\mu$m and 1\,m.
For the 1\,mbar calculation, no condensates are stable for $T\gtrsim2900\,$K, thus the condensation timescale cannot be calculated.
In case of the 100\,bar atmosphere, this threshold is at $T\approx4900\,$K.

The timescale based on reorganisation of condensates in the lattice structure itself  is based on the transfer of sub groups of molecules from one point in the lattice to another by 
solid diffusion.
This annealing process is described by \citet{1996A&A...312..624D} and \citet{1999A&A...347..594G} and the timescale can be calculated by
\begin{align}
\tau_\mathrm{annealing} = \frac{(\Delta a)^2}{\frac{1}{3}\lambda^2\nu \exp \left(\frac{-E_\mathrm{a}}{k_\mathrm{B}T}\right)}.\label{eq:tau_annealing}
\end{align}
$\lambda$ is the step length,
$\nu$ the oscillation frequency,
$E_\mathrm{a}$ the activation energy that needs to be overcome in order to move from one to an adjacent lattice place,
$k_\mathrm{B}$ the Boltzmann constant
and $T$ the temperature.
$\Delta a$ is the total distance that the particle needs to travel in the lattice structures by random walk.

The black lines in Fig.~\ref{fig:timescales_BSE} show the annealing timescales for different annealing distances $\Delta a$ of $1\,\mu$m and 1\,m.
As \citet{1999A&A...347..594G}, we use the typical \ce{SiO4} vibration frequency ${\nu = 2\cdot10^{13}\,\mathrm{s}^{-1}}$ for the annealing process.
\ce{SiO} nucleation experiments from \citet{1982JChPh..77.2639N} resulted in a characteristic activation energies for silicates of ${E_\mathrm{a}/k_\mathrm{B} = 41000\,}$K.
We assume a step size of $\lambda = 1\,\mathrm{nm}$, based on the order of magnitude of the monomer size calculated by $\lambda = \sqrt[3]{V_1}$ for monomer volumes $V_1$.

For the formation of phyllosilicates, the chemisorption of \ce{H2O} in silicates is of importance \citep{Thi2018}.
The formula to calculate these timescales is also given by Eq.~\ref{eq:tau_annealing}.
\cite{Thi2018} calculate the oscillation frequency for \ce{H2O} a surface sight as $\nu\,=\,10^{12}\,\mathrm{s}^{-1}$.
The activation energy for a chemisorbed \ce{H2O} at the surface to occupy a free silicate core chemisorption site is $E_\mathrm{a}/k_\mathrm{B} = 13470\,$K \citep{Okumura2004}.
The chemisorption timescales for different thicknesses of 1\,m and $1\,\mu$m are shown in Fig.~\ref{fig:timescales_BSE} as blue lines.

The timescales in Fig.~\ref{fig:timescales_BSE} show that the formation of new condensates from the gas phase is quicker than 1\,yr for all temperatures and pressures considered. 
However, the formation of the thermodynamically most favourable lattice structure by rearrangement via solid diffusion may take a long time.
For $\mu$m-sized particles, the silicate rearrangement (annealing) timescale exceeds 1\,Myr at about 900\,K.
In comparison, the chemisorption timescale for the diffusion of water into the rock structure is faster, exceeding 1\,Myr at about 300\,K for $\mu$m-sized particles.

The comparison of condensation timescales and annealing timescales provides a first order insight to the structure of the condensates.
The condensate will be crystalline if ${\tau_\mathrm{cond} > \tau_\mathrm{annealing}}$, because the condensates can rearrange quick enough to form the crystalline structure.
For ${\tau_\mathrm{cond} \ll \tau_\mathrm{annealing}}$ the condensation occurs much faster and the formed condensate is amorphous \citep{1999A&A...347..594G}.

Especially the timescales for the phyllosilicate formation are very interesting as small rocks can be hydrated on timescales of planetary evolution which is in agreement to the conclusions of \cite{Thi2018}.
This underlines the importance of the inclusion of phyllosilicates to atmosphere-crust models.

\section{Summary and Discussion}\label{sec:discussion}
The composition of the near-crust atmosphere of a rocky planet depends on the composition of the crust, and vice versa, as the atmosphere and the crust form a coupled thermo-chemical system.
The gas in such near-crust atmospheres will diffuse (or be transported) into the higher, low-pressure atmospheric regions where it may become, for example, subject to cloud-formation and/or mass loss.
In our model we investigate the composition of the atmospheric gas directly in contact with the crust based on chemical phase equilibrium.
The assumed total element abundances, and the thermo-chemical data of the condensed species included, are crucial factors for the determination of the composition of crust and atmosphere.
All temperature values correspond to an atmospheric pressure of 100\,bar.\\*[-8mm]

\paragraph{Near-crust gas composition regimes:}
The near-crust atmospheres of the investigated rocky compositions (CC, BSE, MORB and CI chondrite) fall roughly into three regimes:
\begin{itemize}
\item[(i)] near-crust atmospheres  with $T\!\la\!600\,$K are dominated by \ce{N2} or \ce{CH4}, 
\item[(ii)] near-crust atmospheres  with 
$600\,\rm{K}\!\la\!T\!\la\!3000\,$K are dominated by  \ce{H2O}, \ce{CO2} and \ce{SO2},  but \ce{O2} dominates for a BSE-crust for $T\!\ga\!2000\,$K,
\item[(iii)] near-crust atmospheres  with  $T\!\ga\!3000\,$K consist of a combination of 
\ce{O2}, metal atoms, oxides and hydroxides.
\end{itemize}
The resulting near-crust atmospheric composition at lower temperatures ($T\!\la\!700\,$K) depends on whether or not phyllosilicates are included. 
This can result in a shift from \ce{CH4} as a dominant gas species to \ce{N2}, as the hydrogen is consumed by condensates in the form of phyllosilicates.
The model results for the CI chondrite abundances are the only ones that produces \ce{CO} in gas concentrations higher than 10\%, due to the relatively high C abundance.

The near-crust atmosphere for a rocky surface according to the polluted white dwarfs metal abundance remains inconclusive with respect to P, S and K species as they are not measured in the PWD spectra.
Nevertheless, our results for PWD abundances provide  insight into the atmospheric composition of a potential iron rich planet.
The major species of the corresponding near-crust gas phase would be \ce{CH4} ($T\!\la\!750\,$K), \ce{H2} ($750\,\rm{K}\!\la\!T\!\la\!3300\,$K), \ce{SiO} ($330\,\rm{K}\!\la\!T\!\la\!4700\,$K) and \ce{Fe} ($4700\,\rm{K}\!\la\!T$).\\*[-8mm]

\paragraph{Implications on magma oceans:}
We observe in our models that liquids start to play a role at temperatures $T\!\ga\!1700\,$K, where parts of the crust are molten for all total element abundances considered. 
For hot super earth planets like 55\,Cnc\,e or CoRoT-7b with dayside temperatures in the range from about $2500\,$K to $3000\,$K the crust is mainly liquid, but also contains some solids.
The crust composition can be grouped into two regimes correlating with the difference in the Mg/Si ratio
\begin{itemize}
\item[(i)] BSE/CI: melt consisting of mainly \ce{MgSiO3}[l], \ce{FeO}[l] and \ce{MgSiO4}[l] while the main solids are \ce{Ca2MgSi2O7}[s] (\aa{\sl kermanite}), \ce{MgCr2O4}[s] ({\sl picrochomite}) and \ce{MnO}[s] ({\sl manganosite}).
\item[(ii)] CC/MORB: melt consisting of mainly \ce{SiO2}[l], \ce{MgTi2O5}[l] and \ce{FeO}[l] while the solids contain \ce{MnSiO4}[s] ({\sl tephroite}), \ce{Cr2O3}[s] ({\sl eskolaite})  and \ce{CaSiO3}[s] ({\sl wollastonite}).
\end{itemize}
The near-crust atmosphere above a magma ocean consists, in all cases, mainly of \ce{H2O}, \ce{O2}, \ce{CO2} and \ce{SO2}. 
The major differences are the dominance of \ce{O2} over \ce{H2O} for the BSE abundances and the occurrence of \ce{H2} and \ce{CO} as the second and third most abundant gases in case of the CI chondrite element abundances.\\*[-8mm]

\paragraph{Water stability:}
Phyllosilicates inhibit the thermal stability of liquid and solid water.
The CI chondrite total element abundance is the only case for which the thermal stability of water as a condensate is not inhibited by phyllosilicates.
In fact, water is abundant enough to overcome the stability of phyllosilicates in the CI case.
Similarly, we are able to force the BSE composition to have liquid water as a thermally stable condensate by adding 12\,wt\% \ce{H2O} to the total element abundance.
An additional 3\,wt\% \ce{H2O} allow solid water to be thermally stable.
This shows the importance of phyllosilicates for the search for planets with potential liquid water.
The role of phyllosilicates for the water content of exoplanets is further emphasised by their continuous formation on Earth's surface on short timescales.
On the other hand, it is hypothesised, that most of the Earth's water is trapped in phyllosilicates in the wet mantle transition zone \citep{2019E&PSL.505...42W}. 
But this zone is not saturated with water and is believed to take up all of Earth's water.

Timescale investigations show that hydration of small rocks occurs on evolutionary timescales also for temperatures as low as 300\,K underlining the importance of phyllosilicates to the system.
 The rearrangement in refractory condensates towards the 
thermodynamically most favourable state is only valid for small scales and temperatures above 900\,K.
However, the major changes of the atmospheric composition at temperatures below 1000\,K are caused by the formation of phyllosilicates and ices. \\*[-8mm]

\paragraph{Implication for habitability:}
The search for habitable and inhabited planets is one of astrophysics ultimate goals.
The occurrence of life will alter the atmosphere and produce some gas species that can be used as biotracers. 
Earlier studies suggest that \ce{O2} and \ce{O3} would be a good biosignature, but more recent studies show that also abiotic sources can produce sufficient levels of \ce{O2} in atmospheres \citep{2014ApJ...792...90D, 2015ApJ...812..137H, ast.2014.1231, 2018ApJ...866...56H}.
This underlines the necessity of understanding the difference between biosignatures and false positives.
Previous studies show that various gas species can be used as biotracers (\ce{O2}, \ce{O3}, \ce{N2O}, \ce{CH4}, \ce{CH3Cl}, \ce{NH3}, \ce{C2H6} and sulphur hazes \citep[][and references therein]{DG-2011, 2019arXiv190411716L}).
In this study we find that some of these species are consistent with equilibrium models, but the occurrence of multiple of these species can be a sign for non equilibrium chemistry.
While \ce{O2} is a major constituent of the atmosphere at $T\!<\!2000\,$K, we are unable to produce stable \ce{O2} at lower temperatures ($T\!\la\!1500\,$K).
Some further gas species that might be linked to life, especially \ce{N2} and \ce{CH4}, only occur at low temperatures.
According to our model and the element abundances analysed in this work, we conclude that the following gas compositions as possible signs for non-equilibrium and potential signature of life, ordered by the strength of argument
\begin{itemize}
\item[(i)] \ce{O2} as abundant gas species for $T\!\la\!1500\,$K,
\item[(ii)] \ce{N2} and \ce{O2} at the same time for $T\!\la\!1500\,$K \citep[see also][]{2016AsBio..16..949S},
\item[(iii)] \ce{NO} or \ce{NO2} for $T\!\la\!1500\,$K,
\item[(iv)] \ce{CH4} and \ce{O2} at the same time.
\end{itemize}
The simultaneous occurrence of \ce{N2} and \ce{O2} with high concentrations is also possible for sufficiently high N abundances.
\ce{N2} is usually outgassing as a volatile whereas \ce{O2} is produced from molten rock at high temperatures, but if we simply assume a much higher N-abundance, both \ce{O2} and \ce{N2} can coexist also at high temperatures in equilibrium.
We note, however, that the detection of \ce{N2} can be affected by lightning in cloudy atmospheres  \cite{2017MNRAS.470..187A}.

One of the crucial aspects in the formation of life as we know it is the occurrence of liquid water.
From the results of this work, the upper crust of a (cooling) planet needs to be saturated in phyllosilicates in order to allow for the stability of liquid water on the surface.
The detection of gaseous water in the atmosphere is not conclusive for the existence of liquid water, as the phyllosilicates are able to incorporate all potential liquid water in phase equilibrium.

Other conclusions about the crust composition on the basis of observations of the atmospheric composition can be ambiguous as well, but can still provide the first steps to characterise the conditions on the surface of terrestrial exoplanets.

\begin{acknowledgements}
O.H. acknowledges the PhD stipend form the University of St Andrews' Centre for Exoplanet Science. We thank Stephen J.~Mojzsis, Inga Kamp, Sami Mikhail and Mark Claire for valuable discussions on phyllosilicates and their stability.
\end{acknowledgements}

\bibliography{bibli} 
\onecolumn{
\begin{appendix}
\section{Additional Figures and Tables}

\begin{table*}[h]
\caption{Glossary for different condensed species.}
\label{tab:glossary}
\centering
\vspace*{-2mm}
\begin{tabular}{ccc}
\hline
 Name & Sum formula& Structural formula\\ \hline
 Dolomite   &   \ce{CaMgC2O6}       &  \ce{CaMg(CO3)2}\\
 Grossular  &   \ce{Ca3Al2Si3O12}   &  \ce{Ca3Al2(SiO4)3}\\
 Phlogopite &   \ce{KMg3AlSi3O12H2} &  
  \multirow{2}{*}{$\biggr\rbrace\;\ce{KMg3(AlSi3O10)(F,OH)2}$} \\
 Flourophlogopite &  \ce{KMg3AlSi3O10F2}   &  \\
 Sodaphlogopite &   \ce{NaMg3AlSi3O12H2}   &  \ce{NaMg3(AlSi3O10)(OH)2}\\
 Talc  &   \ce{Mg3Si4O12H2}   &   \ce{Mg3Si4O10(OH)2}\\
 Fe-Chloritoid  &  \ce{FeAl2SiO7H2}   &   
  \multirow{3}{*}{$\Biggr\rbrace\;\ce{(Fe,Mg,Mn)2Al4Si2O10(OH)4}$} \\
 Mg-Chloritoid  &  \ce{MgAl2SiO7H2}   &   \\
 Mn-Chloritoid  &  \ce{MnAl2SiO7H2}   &   \\
 Lizardite  &   \ce{Mg3Si2O9H4}   &   \ce{Mg3Si2O5(OH)4}\\
 Greenalite &   \ce{Fe3Si2O9H4}   &   \ce{Fe3Si2O5(OH)4}\\
 Goethite   &   \ce{FeO2H}  &   \ce{FeO(OH)}\\
 Epidote    &   \ce{Ca2FeAl2Si3O13H}    &   \ce{Ca2(FeAl2)(Si2O7)(SiO4)O(OH)}\\
 Lawsonite  &   \ce{CaAl2Si2O10H4}  &   \ce{CaAl2(Si2O7)(OH)2(H2O)}\\
 Prehnite   &   \ce{Ca2Al2Si3O12H2} &   \ce{Ca2Al2Si3O10(OH)2}\\
 Ferri-Prehnite &  \ce{Ca2FeAlSi3O12H2}   &   \ce{Ca2FeAlSi3O10(OH)2}\\
 Margarite  &   \ce{CaAl4Si2O12H2}   &   \ce{CaAl2Al2Si2O10(OH)2}\\ \hline
\end{tabular}
\end{table*}

\begin{figure*}[!h]
\centering
\vspace*{-2mm}
\includegraphics[width = .490\linewidth, page=8]{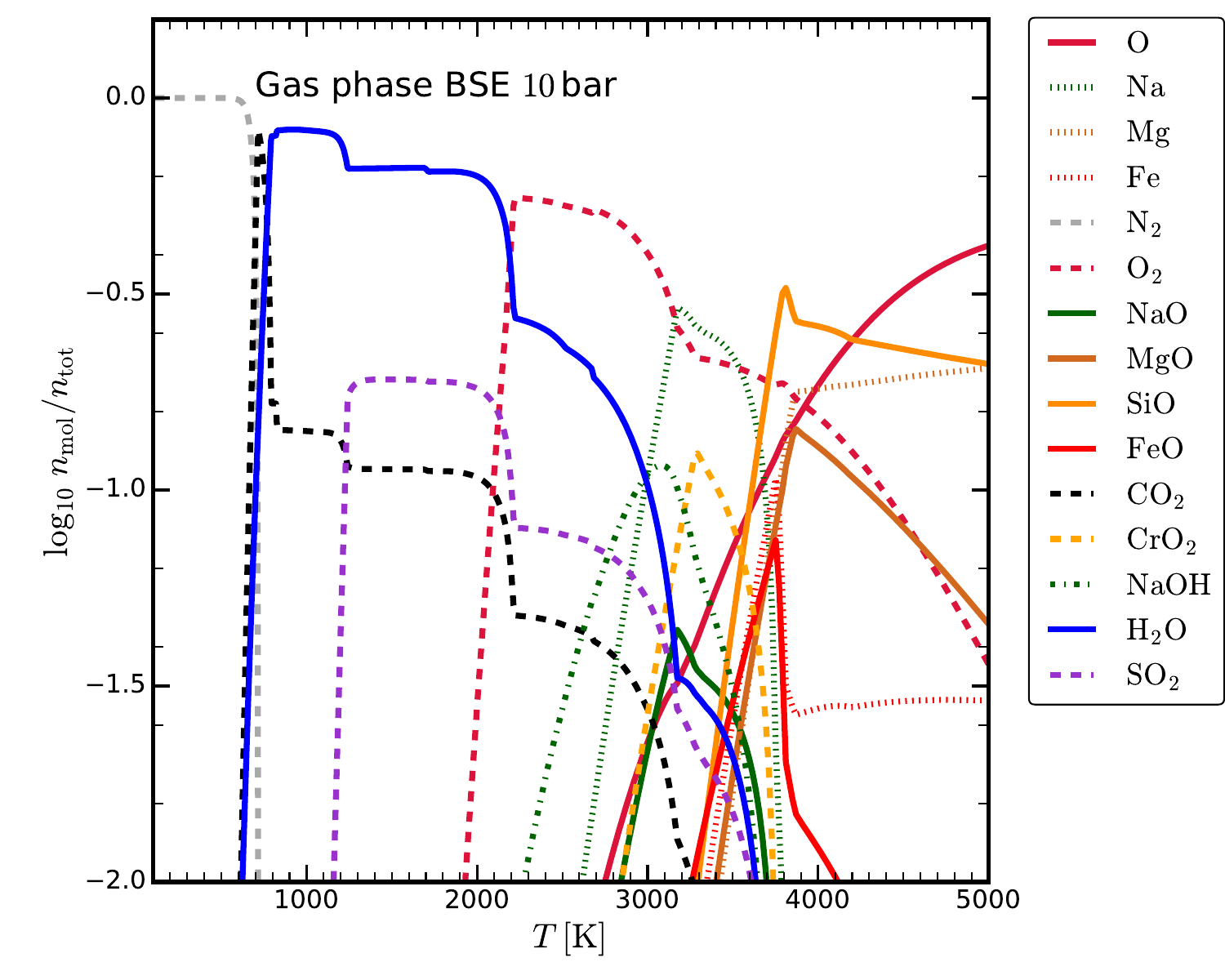}
\includegraphics[width = .490\linewidth, page=8]{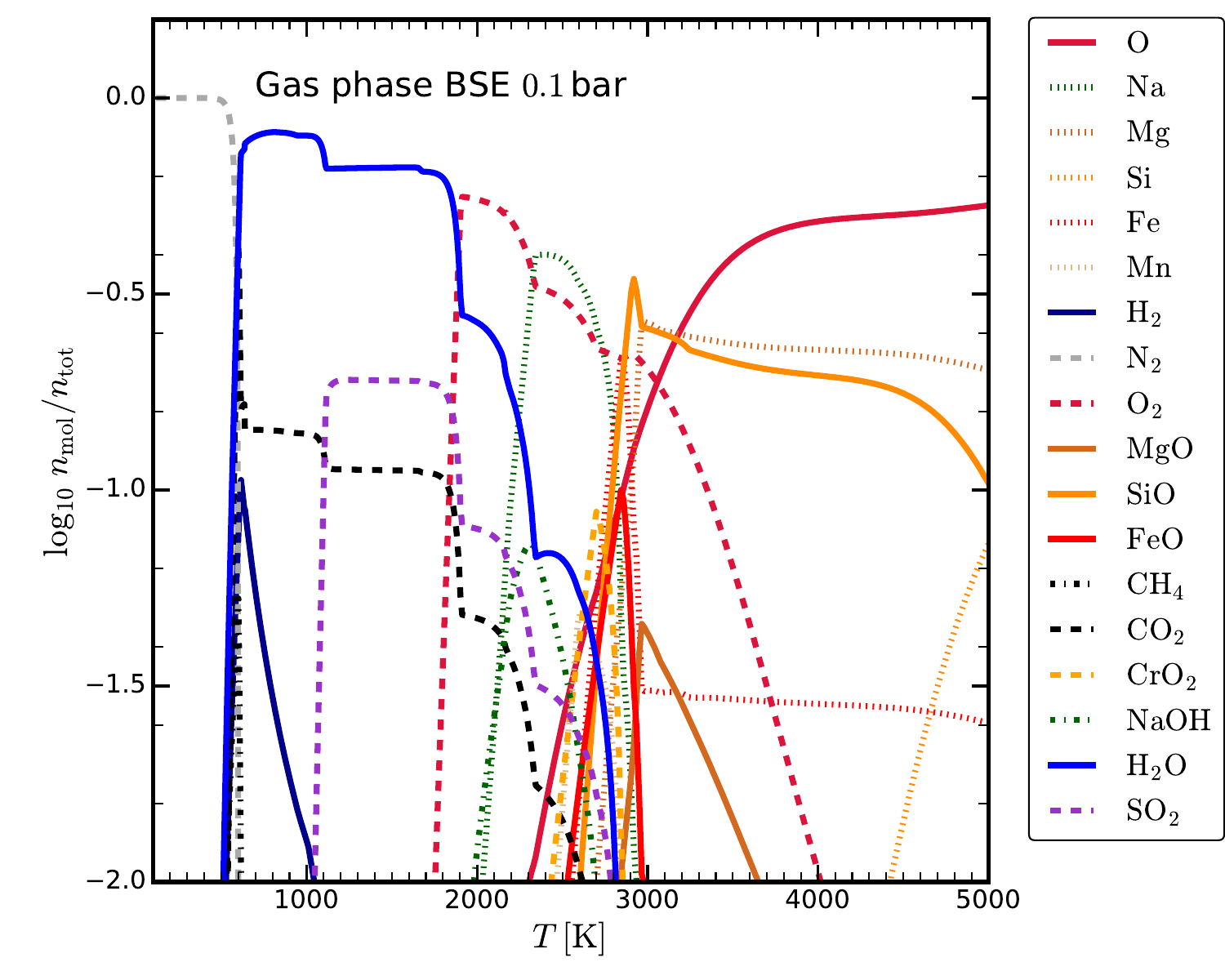}
\caption{Two further pressure profiles for the pressure levels of 10 bar and 0.1 bar are shown as described in figure \ref{fig:p-T_plane_BSE}.}
\label{fig:p-T_plane_BSE_2}
\end{figure*}

\begin{table*}[h]
\caption{Element abundances from different astronomical and geological sources. "H normalised" means $\log_{10}$ of nuclei particle ratio with respect to hydrogen, where H is nomalised to 12, and "mfrac [\%]" means mass fraction in the condensate.
References.}
\label{tab:abundances}
\centering
\vspace*{-2mm}
\begin{tabular}{c|cccccc}
\hline
    & & & & & & \\[-2.2ex] 
Element	&	Solar	&	\!\!CC Schaefer	&	\!\!BSE Schaefer	&	MORB	&	CI meteorite	&	Polluted White dwarf	\\
\hline
    & & & & & & \\[-2.2ex] 
Refernces    & 1
	& 2& 2
	& 3
	& 4
	& 5 \\
	&	H normalised &	mfrac [\%]	&	mfrac [\%]	&	mfrac [\%]	&	mfrac[\%]	& H normalised 	\\ \hline
	& & & & & & \\[-2.2ex]
H	&	12	         &  0.045	    &	0.006	    &	0.045	    &	1.97	    & 12 \\
He	&		         &		        &		        &	0.000006	&	0.000000917	& 13.14 \\
Li	&		         &	      	    &		        &	0.00065	    &	0.000147	& \\
Be	&		         &		        &		        &	0.000076	&	0.0000021	& \\
B	&		         &		        &		        &	0.00016	    &	0.0000775	& \\
C	&	8.5695	     &	0.199	    &	0.006	    &	1.5	        &	3.48	    & <\,7.5	\\
N	&	8.0704	     &	0.006	    &	8.8E-05	    &	0.14	    &	0.295	    & <\,9	\\
O	&	8.83	     &	47.2	    &	44.42	    &	43.94564	&	45.9	    & <\,8	\\
F	&	4.8867	     &	0.053	    &	0.002	    &	0.0087	    &	0.00582	    &	\\
Ne	&		         &		        &		        &		        &	0.000000018	&		\\
Na	&	6.2757	     &	2.36	    &	0.29	    &	2.06978	    &	0.499	    &	$5.35\pm0.20$	\\
Mg	&	7.5233	     &	2.2	        &	22.01	    &	4.57101	    &	9.68	    &	$7.16\pm0.25$	\\
Al	&	6.427	     &	7.96	    &	2.12	    &	7.77999	    &	0.85	    &	$6.74\pm0.20$	\\
Si	&	7.4976	     &	28.8	    &	21.61	    &	23.59144	&	10.7	    &	$7.3\pm0.30$	\\
P	&	5.4798	     &	0.076	    &	0.008	    &	0.08030	    &	0.0967	    &		\\
S	&	7.1965	     &	0.07	    &	0.027	    &	4	        &	5.35	    &		\\
Cl	&	5.2534	     &	0.047	    &	0.004	    &	0.037	    &	0.0698	    &		\\
Ar	&		         &		        &		        &	    	    &	0.000000133	&		\\
K	&	5.1208	     &	2.14	    &	0.02	    &	0.13282	    &	0.0544	    &		\\
Ca	&	6.3555	     &	3.85	    &	2.46	    &	8.14033	    &	0.922	    &	$6.9\pm0.10$	\\
Sc	&	3.0416	     &		        &		        &	0.00398	    &	0.00059	    &	$2.95\pm0.30$	\\
Ti	&	4.9408	     &	0.401	    &	0.12	    &	1.00690	    &	0.0451	    &	$5.19\pm0.10$	\\
V	&	3.9159	     &		        &		        &	0.0309	    &	0.00543	    &	$4.4\pm0.30$	\\
Cr	&	5.6014	     &	0.013	    &	0.29	    &	0.0249	    &	0.265	    &	$5.73\pm0.10$	\\
Mn	&	5.4661	     &	0.072	    &	0.11	    &	0.14250	    &	0.193	    &	$5.67\pm0.10$	\\
Fe	&	7.4167	     &	4.32	    &	6.27	    &	8.10729	    &	18.5	    &	$7.49\pm0.10$	\\
Co	&	6.1788	     &		        &		        &	0.0043	    &	0.0506	    &	$4.64\pm0.40$	\\
Ni	&		         &		        &		        &	0.0092	    &	1.08	    &	$6.07\pm0.15$	\\
Cu	&	4.23	     &		        &		        &	0.0074	    &	0.0131	    &	$3.94\pm0.40$	\\
Zn	&		         &		        &		        &	0.00913	    &	0.0323	    &		\\
Ga	&		         &		        &		        &	0.00175	    &	0.000971	&		\\
Ge	&		         &		        &		        &	0.0021	    &	0.00326	    &		\\
As	&		         &		        &		        &	0.00018	    &	0.000174	&		\\
Se	&		         &		        &		        &	0.0013	    &	0.00203	    &		\\
Br	&	2.6279	     &		        &		        &	0.00012	    &	0.000326	&		\\
Kr	&	    	     &	            &		        &		        &	0.00000000522 &		\\
Rb	&		         &		        &		        &	0.000288	&	0.000231	&		\\
Sr	&	2.9273	     &		        &		        &	0.0129	    &	0.000781	&	$2.72\pm0.30$	\\
Y	&	2.1788	     &		        &		        &	0.00368	    &	0.000253	&		\\
Zr	&	2.9447	     &		        &		        &	0.01169	    &	0.000362	&		\\
W	&	 	         &		        &		        &	0.000012	&	0.0000096	&		\\ \hline
    & & & & & & \\[-2.2ex] 
Sum	&	             &	99.812		&  99.773	    &	105.423	    &	100.066  	&	\\ \hline
\end{tabular}
\tablebib{(1)~\citet{1990ApJS...72..417S}; (2)~\cite{2012ApJ...755...41S}; (3)~\cite{2013GGG....14..489G}; (4)~\cite{2009LanB...4B..712L}; (5)~\cite{2007ApJ...671..872Z}}
\end{table*}

\begin{table*}[h]
\caption{Temperatures where different condensed species appear ($T_{\rm{high}}$) and disappear ($T_{\rm{low}}$) in phase equilibrium for BSE abundances. Three different pressure levels are considered as indicated. The condensates are ordered by the temperature of their first appearance at $p\!=\!100$\,bar.}
\label{tab:condensates}
\centering
\begin{tabular}{cc|cc|cc|cc}
\hline
& & & & & & & \\[-2.2ex] 
Condensates	&	Name	&	$T_{\rm{high}}$\,[K]	&	$T_{\rm{low}}$\,[K] &	$T_{\rm{high}}$\,[K]	&	$T_{\rm{low}}$\,[K] &	$T_{\rm{high}}$\,[K]	&	$T_{\rm{low}}$\,[K]	\\
& & \multicolumn{2}{c|}{$p\!=\!100$\,bar} &\multicolumn{2}{c|}{$p\!=\!1$\,bar}	&\multicolumn{2}{c|}{$p\!=\!1$\,mbar}	\\	\hline
& & & & & & & \\[-2.2ex]   
\ce{MgAl2O4} [l]	&		&	4871	&	2403	&	3653	&	2403	&	2532	&	2403	\\
\ce{MgO}[l]	&		&	4562	&	3849	&		&		&		&		\\
\ce{CaO} [l]	&		&	4562	&	3799	&		&		&		&		\\
\ce{MgSiO3} [l]	&		&	4387	&	1850	&	3290	&	1850	&	2341	&	1850	\\
\ce{FeO} [l]	&		&	4387	&	1644	&	3247	&	1644	&	2281	&	1644	\\
\ce{MgTiO3} [l]	&		&	4330	&	2499	&	3205	&	2499	&		&		\\
\ce{Mg2SiO4} [l]	&		&	3849	&	2164	&	3333	&	2164	&	2403	&	2164	\\
\ce{MnO}	&	Manganosite	&	3849	&	2341	&	2924	&	2341	&		&		\\
\ce{Na2SiO3} [l]	&		&	3799	&	2193	&	2703	&	2193	&		&		\\
\ce{Ca2SiO4}	&	Larnite	&	3799	&	3205	&	3421	&	3205	&		&		\\
\ce{Cr2O3}	&	Eskolaite	&	3701	&	3701	&		&		&		&		\\
\ce{Mg3P2O8}	&		&	3512	&	3247	&		&		&		&		\\
\ce{MgCr2O4}	&	Picrochomite	&	3377	&	320	&	2963	&	320	&	2251	&	320	\\
\ce{Ca5P3O13H}	&	Hydroxyaoatite	&	3247	&	3081	&	2532	&	2372	&		&		\\
\ce{Ca2MgSi2O7}	&	\AA kermanite	&	3205	&	1975	&	3205	&	1975	&	2403	&	1975	\\
\ce{P2O10}	&		&	3122	&	1688	&	2372	&	1666	&	1850	&	1581	\\
\ce{K2SiO3} [l]	&		&	2886	&	2532	&		&		&		&		\\
\ce{Al2O3} [l]	&		&		&		&		&		&	2668	&	2532	\\
\ce{Ca2Al2SiO7}	&	Gehlenite	&		&		&		&		&	2532	&	2435	\\
\ce{KAlSiO4}	&	Kalsilite	&	2532	&	2081	&	2341	&	2081	&		&		\\
\ce{MgTi2O5} [l]	&		&	2499	&	2001	&	2499	&	2001	&	2311	&	2001	\\
\ce{Ca3MgSi2O8}	&	Merwinite	&		&		&		&		&	2435	&	2403	\\
\ce{MgAl2O4}	&	Spinel	&	2403	&	2136	&	2403	&	2136	&	2403	&	2136	\\
\ce{MgFe2O4}	&	Magnesioferrite	&	2372	&	1924	&	2027	&	1924	&		&		\\
\ce{MnTiO3} 	&	Pyrophanite	&	2341	&	1850	&	2341	&	1850	&	2136	&	1850	\\
\ce{NaAlSiO4}	&	Nepheline	&	2193	&	1949	&	2193	&	1949	&	380	&	277	\\
\ce{Mg2SiO4}	&	Forsterite	&	2164	&		&	2164	&		&	2164	&		\\
\ce{CaAl2Si2O8}	&	Anorthite	&	2136	&	570	&	2136	&	570	&	2136	&	570	\\
\ce{KAlSi2O6}	&	Leucite	&	2081	&	1155	&	2081	&	1155	&	1874	&	1155	\\
\ce{Fe2TiO4}	&	Ulvöspinel	&	2001	&	1233	&	2001	&	1233	&	2001	&	1233	\\
\ce{CaMgSi2O6}	&	Diopside	&	1975	&	347	&	1975	&	347	&	1975	&	347	\\
\ce{NaAlSi3O8}	&	Albite	&	1949	&	380	&	1949	&	380	&	1924	&	380	\\
\ce{Fe3O4}	&	Magnetite	&	1924	&		&	1924	&		&	1733	&		\\
\ce{SiO2} [l]	&		&	1850	&	1826	&	1850	&	1826	&	1850	&	1826	\\
\ce{MnSiO3}	&	Pyroxmangite	&	1850	&	1581	&	1850	&	1581	&	1850	&	1581	\\
\ce{MgSiO3}	&	Enstatite	&	1826	&		&	1826	&		&	1826	&		\\
\ce{Ca5P3O12F}	&	Flourapatite	&	1710	&		&		&		&	1623	&		\\
\ce{FeO}	&	Ferropericlase	&	1644	&	1000	&	1644	&	1000	&	1644	&	1000	\\
\ce{Mn2SiO4}	&	Tephroite	&	1581	&	1560	&	1581	&	1560	&	1581	&	1560	\\
\ce{MnTiO3} *	&	Pyrophanite	&	1560	&	1170	&	1560	&	1170	&	1560	&	1170	\\
\ce{NaCl} [l]	&		&	1462	&	1068	&		&		&		&		\\
\ce{FeS}	&	Troilite	&	1316	&		&	1170	&		&	1000	&		\\
\ce{FeTiO3}	&	Ilmenite	&	1233	&		&	1233	&		&	1233	&		\\
\ce{Mn3Al2Si3O12}	&	Spessartine	&	1170	&		&	1170	&		&	1170	&		\\
\ce{KAlSi3O8}	&	Microcline	&	1155	&	962	&	1155	&	712	&	1155	&	513	\\
\ce{KMg3AlSi3O10F2}	&	Flourophlogopite	&	1140	&	427	&	925	&	427	&	731	&	427	\\
\ce{NaCl}	&	Halite	&	1068	&		&	1054	&		&	780	&		\\
\ce{Fe2SiO4}	&	Fayalite	&	1000	&	150	&	1000	&	150	&	1000	&	150	\\
\ce{KMg3AlSi3O12H2}	&	Phlogopite	&	962	&		&	712	&		&	513	&		\\
\ce{NaMg3AlSi3O12H2}	&	Sodaphlogopite	&	925	&		&	684	&		&	493	&		\\
\ce{C}	&	Graphite	&	780	&		&	649	&		&	487	&		\\
\ce{FeAl2O4}	&	Hercynite	&	570	&		&	570	&		&	570	&		\\
\ce{MgF2}	&		&	427	&	187	&	427	&	187	&	427	&	187	\\
\ce{NaAlSiO4} *	&	Nepheline	&	380	&	277	&	380	&	277	&	380	&	277	\\
\ce{Ca3Al2Si3O12}	&	Grossular	&	347	&		&	347	&		&	347	&		\\
\ce{Cr2O3} *	&	Eskolaite	&	320	&		&	320	&		&	320	&		\\
\ce{NaAlSi2O6}	&	Jadeite	&	277	&		&	277	&		&	277	&		\\
\ce{CaF2}	&		&	187	&	100	&	187	&	100	&	187	&	100	\\
\ce{Fe}	&	Iron	&	150	&		&	150	&		&	150	&		\\
\ce{MgF2} *	&		&	100	&		&	100	&		&	100	&		\\
\hline
\end{tabular}\\
The * indicates species, that are stable at a second temperature range.
\end{table*}

\begin{figure}[!t]
\hspace*{-4mm}
\includegraphics[width = 94mm, page=8] 
  {./graphs/BSE_water/ggchem_100+1Fe.pdf}
\hspace*{-1mm}
\includegraphics[width = 94mm] 
  {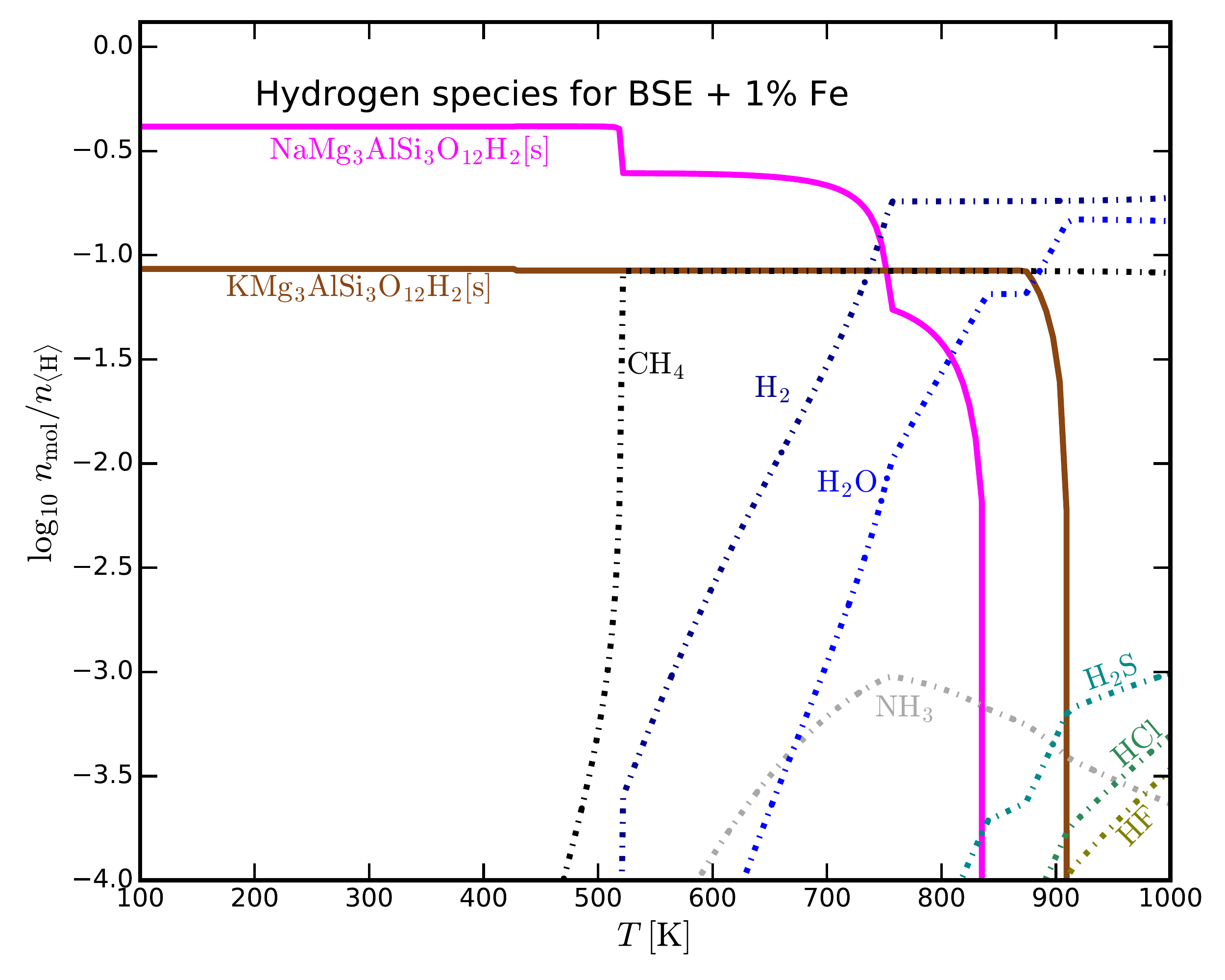}\\[-4ex]
\caption{Results for {\sl BSE} total element abundances at constant pressure $p\!=\!100$\,bar. Phyllosilicates are included as possible condensates.
{\bf Left panel:} gas phase concentrations ($n_\mathrm{mol}/n_\mathrm{tot}$) between 100\,K and 5000\,K. All species with maximum $\log n_\mathrm{mol}/n_\mathrm{tot}\! >\!-1.4$ are shown. {\bf Right panel:} gaseous and condensed species that contain hydrogen are plotted with their concentration per H-nucleus $n/n_{\langle H\rangle}$ between 100\,K and 1000\,K (note the different scalings).
}
\label{fig:GGchem-BSE+Fe}
\end{figure}
\end{appendix}
}
\end{document}